\documentclass[12pt,titlepage]{utarticle}
\usepackage{amssymb,color}
\usepackage{hyperref}

\numberwithin{equation}{section}

\def\del{\partial}
\def\delb{{\bar{\partial}}}
\newcommand{\vvev}[1]{{\langle\kern-.5ex\langle #1\rangle\kern-.5ex\rangle}}

\newcommand{\tr}[1]{{\mbox{tr}\left({#1}\right)}}          

\newcommand{\mf}[1]{{{\bf {#1}}}}
\def\Z{\mathbb{Z}}
\def\C{\mathbb{C}}

\def\Q{\mathbb{Q}}
\def\I{\mathbf{I}}

\renewcommand{\CP}[1]{{\mathbb{C}{\mathbf{P}^{#1}}}}        

\def\mcO{\mathcal{O}}

\def\mcM{\mathcal{M}}

\def\Tr{\mathrm{Tr}}

\def\N{\mathcal{N}}

\def\<{\langle}
\def\>{\rangle}
\def\P{{\Phi}}
\def\Pb{{\bar{\Phi}}}
\def\Db{{\bar{D}}}
\def\Wb{{\bar{W}}}
\def\Ab{{\bar{A}}}

\def\Ib{{\bar{\mathbf{I}}}}

\def\ib{{\bar{\imath}}}
\def\jb{{\bar{\jmath}}}
\def\kb{{\bar{k}}}
\def\lb{{\bar{l}}}
\def\nb{{\bar{n}}}
\def\a{{\alpha}}
\def\d{{\delta}}

\def\p{{\phi}}
\def\L{{\mathcal{L}}}

\begin{document}
\preprint{UTTG--07--05\\
\texttt{hep-th/0507090}\\}

\title{Random Polynomials and the Friendly Landscape}

\author{Jacques Distler and Uday Varadarajan}

\oneaddress{Theory Group, Physics Department\\
             University of Texas at Austin\\
             Austin, TX 78712\\ {~}\\
             \email{distler@golem.ph.utexas.edu}
             \email{udayv@physics.utexas.edu}}

\Abstract{  
  In hep-th/0501082, a field theoretic ``toy model'' for the Landscape
  was proposed. We show that the considerations of that paper carry
  through to realistic effective Lagrangians, such as those that
  emerge out of string theory. Extracting the physics of the large
  number of metastable vacua that ensue requires somewhat more
  sophisticated algebro-geometric techniques, which we review.}

\maketitle
\newpage

\section{Introduction and Summary}\label{sec:intro}

One of the striking observations of KKLT \cite{Kachru:2003aw} was that
the existence of a large number of long-lived metastable vacua in
certain compactifications of string theory (dubbed the ``landscape''
\cite{Susskind:2003kw}) provides a concrete instance in which the
anthropic principle \cite{Weinberg:1987dv, Brown:1987dd,
  Weinberg:1988cp, Martel:1997vi, Garriga:1999bf, Weinberg:2000qm,
  Bousso:2000xa} might be realized in Nature.  Unfortunately, many
discussions of these ideas get bogged down because it is hard to
disentangle the intricacies of the string theoretic constructions from
the ``anthropic'' questions one would like to address.

The paper of Arkani-Hamed, Dimopoulos and Kachru
\cite{Arkani-Hamed:2005yv} was, therefore, very useful in clearing
away the string-theoretic underbrush and presenting a simple,
tractable field-theoretic model with a large number of vacua in which
anthropic questions could be addressed (see also
\cite{Dienes:2004pi}).

One of the concepts to emerge from their investigation was the notion
of a ``friendly landscape.''  In general, all of the couplings,
$c_{a}$ of the theory will vary between the different vacua. However,
there is a qualitative difference between those couplings which
``scan'' (those whose standard deviation is much larger than their
mean value) and those couplings which ``don't scan'' (those which are
sharply-peaked about their mean value). In making anthropic arguments,
one usually considers the situation in which one coupling is allowed to
vary, while the others are held fixed. If all the couplings vary
appreciably, the anthropic bounds are much weaker, or go away
entirely. So it was very useful for the authors of
\cite{Arkani-Hamed:2005yv} to identify a \emph{mechanism} by which the
couplings one would like to ``tune'' anthropically \emph{scan},
whereas the remaining couplings are sharply peaked.

Their model had three basic features:
\begin{itemize}
\item[1)]A large number, $N$, of scalar fields, $\phi_{i}$.
\item[2)]A decoupled  form for the scalar potential,
  $V(\phi)=\sum_{i}V_{i}(\phi_{i})$.
\item[3)]A decoupled form for the $\phi$-dependence of the observable
  couplings of the model, $c_{a}(\phi)=\sum_{i}c_{ai}(\phi_{i})$.
\end{itemize}
The first has a fairly natural realization in string theory. Many
compactification of string theory have hundreds of moduli which, when
the physics which lifts the vacuum degeneracy is included \cite{
  Giddings:2001yu, Kachru:2002he, Acharya:2002kv, Acharya:2003gb,
  Giryavets:2003vd, Denef:2004dm, Denef:2005mm,
  Balasubramanian:2004uy, Balasubramanian:2005zx, Conlon:2005ki,
  DeWolfe:2005uu}, form natural candidates for the $\phi_{i}$. We can,
quite plausibly, take ``$N$'' to be the number of (complex) moduli.
However, 2,3) are rather unnatural from this point of view and,
indeed, it's hard to imagine that such a decoupled structure might
emerge from string theory.

In the present work, we would like to overcome this drawback and
present what we hope is a realistic version of the scenario of
\cite{Arkani-Hamed:2005yv}. Our approach will be to start with,
essentially, the most general $\N=1$ supersymmetric effective field
theory with a large number of ``moduli'' chiral multiplets,
$\Phi^{i}$. We will then see what conditions must be imposed in order
to realize a friendly landscape.

In \S\ref{sec:sectors}, we will discuss the theory of $N$ chiral
multiplets coupled to $\N=1$ supergravity. Whereas any such theory
must be cut off at a scale $M_{c}< M_{p}$, we will see that, at
large-$N$, we will will need to impose the stronger condition $M_{c}<
M_{p}/\sqrt{N}$, in order to have a sensible effective field theory.
Moreover, we will find, in \S\ref{subsec:local}, that the couplings
among the moduli chiral multiplets must be suitably small; we will
summarize the condition that we must impose on these couplings by
saying that they are \emph{generically small}. We then turn to a
polynomial truncation of the superpotential of the model. While not
indispensable, such a truncation makes the analysis of the vacuum
structure amenable to perturbation theory. As usual, for this to be a
\emph{reliable} guide to the vacuum structure, the super-renormalizable
terms in the superpotential must have coefficients governed by a mass
scale $M_r \ll M_{c}$ (\S\ref{subsec:VacStruct}). In \S\ref{subsec:Flux},
we will see how these considerations mesh with the most popular arena
for landscape considerations --- F-theory vacua with fluxes. 

Next (\S\ref{subsec:R}), we impose the discrete R-symmetry found by
\cite{Arkani-Hamed:2005yv} to lead to a friendly landscape for the
cosmological constant, and use the $GL(N,\BC)$ symmetry of field
redefinitions to simplify our problem (\S\ref{subsec:Fermat}).

In \S\ref{sec:Solve}, we lay out the algebro-geometric techniques used
to determine the vacuum structure and extract information about the
distribution of values for various physical quantities among the
$2^{N}$ vacua of the theory. In \S\ref{sec:Stats}, we use these
techniques to compute certain ``holomorphic moments'' which
characterize the distribution of values of the cosmological constant.
In \S\ref{sec:Scanning}, we discuss the physics that ensues when one
assumes that the couplings are chosen from some (unspecified)
probability distribution, generalizing the considerations of
\cite{Arkani-Hamed:2005yv}. We apply our analysis both to the
superpotential and to other holomorphic couplings.

Finally, in \S\ref{sec:Generalizations}, we discuss some
generalizations of our techniques and future directions.

\section{General Features of SUSY Landscape Sectors}\label{sec:sectors} 

As a field theoretic model for the landscape sector, we could start
with an arbitrary SUSY field theory of $N$ chiral superfields and $N'$
vector superfields, all coupled to $\N=1$ SUGRA. For simplicity, we
will restrict ourselves to the case $N'=0$ and only briefly touch on
generalizations to gauged hidden sectors in this work. Thus, the
vacuum structure of the model can be determined using the
two-derivative effective action of the SUSY non-linear $\sigma$-model
describing the $N$ chiral fields at energies below a cutoff scale,
$M_c$.

Clearly, for any sort of effective field theory to be valid, we must
take $M_{c}<M_{p}$. However, as noted in \cite{Arkani-Hamed:2005yv},
when one has a large number, $N\gg 1$, of fields, radiative stability
of Newton's constant requires
\begin{equation}
   \frac{M_{c}^{2}}{M_{p}^{2}}< \frac{1}{N}
\end{equation}
To see this, note that the action for $\N=1$ supergravity
interacting with $N$ chiral fields is, in superspace notation (we
use $M_p$ for the reduced Planck mass and the conventions of
\cite{Gates:1983nr}),
\begin{equation}
  \label{eq:SUGRAaction}
  S =  -3M_p^2 \int d^8 z E^{-1} e^{- \frac{1}{3M_p^2}K(\P,\Pb)}
    + \int d^6 z \phi^3 W(\P) + \mathrm{h.c.}
\end{equation}
where $E^{-1}$ is the superdeterminant of the vielbein, $z$ is a
superspace coordinate, and $\phi$ is a compensator superfield. The
Einstein-Hilbert term comes from the leading ($\P$-independent) piece
of the first term in \eqref{eq:SUGRAaction}. At one-loop, this
receives a quadratically-divergent contribution,
\begin{equation}
 \left. \Delta K_W^{(1)}\right|_{\P=\Pb=0} \sim
    \left(\frac{M_c^2}{16\pi^2} \right) g^{i \jb} \del_i \delb_\jb K
    \sim N \left(\frac{M_c^2}{16\pi^2} \right) 
\end{equation}
where $g_{i\overline{\jmath}}= \del_{i}\delb_{\jb} K$ is the K\"ahler
metric of the $\sigma$-model.  The enhancement which comes from having
$N$ fields running around the loop requires us to set the cutoff,
$M_{c}$ to be parametrically smaller than $M_{p}$.

In $\N=1$ supergravity, the chiral multiplets parameterize a K\"ahler
manifold, whose K\"ahler form, $\omega = \tfrac{i}{2}
\partial\overline{\partial} K$. The superpotential, $W(\P)$ transforms
as a section of a line bundle, $\CL$, whose first Chern class,
\begin{equation}
c_1(\L) = -\frac{1}{\pi M_{p}^{2}} \omega = \frac{1}{2\pi i M_p^2} \del \delb K.
\end{equation}
The fiber metric on $\CL$ is
\begin{equation}
h(\P,\Pb)=e^{K(\P,\Pb)/M_p^2}
\end{equation}
and the connection is of type (1,0), $D_i = \del_i + \del_i K/M_p^2$.
Under K\"ahler transformations, $W(\P)\to e^{f(\P)/M_{p}^{2}} W(\P)$,
\begin{equation}
  \label{eq:Kpot}
K(\P^i,\Pb^\ib) \rightarrow K(\P^i,\Pb^\ib) + f(\P^i) +
\bar{f}(\Pb^\ib),
\end{equation}
The supersymmetric vacua of this model are the critical points of the
superpotential with respect to the Chern connection, i.e. the points
at which $D_i W \in \Gamma(T^*X \otimes \L)$ intersect the zero
section of $T^*X \otimes \L$,
\begin{equation}
  \label{eq:SUSYvac}
D_i W = \del_i W + (\del_i K) W /M_p^2=0.
\end{equation}
More generally, all the vacua of this model, supersymmetric or not,
are critical points of its scalar potential,
\begin{equation}
  \label{eq:scalarpot}
  V = e^{K/M_p^2} \left( g^{i \jb}  D_i W \Db_\jb \Wb - 3 |W|^2/M_p^2 \right).
\end{equation}
Since we will be interested in vacua in which $|W|\sim M_{r}^{3} <
M_{c}^{3} \ll M_{p}^{3}$, we will always be safe in neglecting the
connection term in \eqref{eq:SUSYvac} and hunting for ordinary
critical points, $\partial_{i}W=0$.

In fact, to justify a perturbative analysis of the effective
Lagrangian \eqref{eq:SUGRAaction}, we require that $W(\P)$ have a
convergent Taylor expansion in a polydisk, $|\P_{i}|< M_{c}$. Just as
the radiative stability of Newton's constant imposed constraints on
the cutoff scale, $M_{c}$ of our effective Lagrangian, radiative
corrections to the K\"ahler metric imposes, at large-N, further
constraints on the coefficients.

\subsection{Quantum Corrections to the K\"{a}hler
  Potential and Large $N$ Scaling}\label{subsec:local}

Before embarking upon a study of the vacua of these models, we should
study the radiative stability of these models in the limit of large
$N$. As our theory is an effective field theory, we expect that the
characteristic radius of curvature for $X$ is given by the cutoff
scale, $M_c$. Thus, in a small enough neighborhood of a smooth point
$p \in X$, we can choose local coordinates $\P^i$ such that
$\P^i(p)=0$ and the K\"{a}hler potential takes the form (modulo
K\"{a}hler transformations),
\begin{equation}
  \label{eq:Kpower}
  K (\P, \Pb) = g_{i \jb} \P_i \Pb_\jb + \Real \sum_{\I_n, \Ib_\nb}
  \frac{1}{n \nb M_c^{n+\nb-2}} K_{\I_n \Ib_\nb} \P^{i_1}
  \cdots \P^{i_n} \Pb^{\ib_1} \cdots \Pb^{\ib_\nb} 
\end{equation}
where the $K_{\I_n \Ib_\nb}$ are dimensionless and symmetric in the
multi-indices $\I_n$ and $\Ib_\nb$ and $g_{i \jb}$ is the K\"{a}hler
metric at $p$. Further, as the superpotential $W$ is holomorphic, it
can be locally expanded as a power series in the holomorphic
coordinates $\P^i$ about the origin $\P^i(p)=0$,
\begin{equation}
  \label{eq:Wexp}
  W(\P^i) = A_0 + \sum_{n, \I_n} \frac{A_{\I_n}}{n} 
  \P^{i_1} \cdots \P^{i_n} = M_c^3 W_0 + \sum_{n, \I_n}
  \frac{W_{\I_n}}{n M_c^{n-3} } \P^{i_1} \cdots \P^{i_n},
\end{equation}
where the $A_{\I_n}$ are symmetric in the multi-indices $\I_n$ and
have mass dimension $3-n$ while the $W_{\I_n} = M_c^{n-3} A_{\I_n}$
are dimensionless.

While the superpotential is holomorphic and not renormalized, the
K\"{a}hler potential is renormalized. It is important to check the
radiative stability of its assumed form given the form of $W$ in the
large $N$ limit.  More precisely, we compute the effective K\"{a}hler
potential (in the Wilsonian sense) at a scale $M_{c'}$ lower than
$M_c$ by integrating out the modes in a shell of momenta between $M_c$
and $M_{c'}$, and require that the corrections are not parametrically
larger in $N$ than the bare values.  As we will see, this requirement
will restrict the asymptotic growth of both the $A_{\I_n}$ and
$K_{\I_n \Ib_\nb}$ parametrically in $N$.

Let us first consider the one-loop corrections to the K\"{a}hler
potential coming from the superpotential $W$,
\begin{equation}
\begin{split}
  \label{eq:WcorrK}
  \Delta K_W^{(1)} & \sim \left(\frac{1}{8\pi^2}
    \log{\frac{M_c^2}{M_{c'}^2}} \right) g^{i
    \ib} g^{j \jb} \del_i \del_j W \delb_\ib \delb_\jb \Wb  \\
  & \sim \left(\frac{1}{8\pi^2} \log{\frac{M_c^2}{M_{c'}^2}} \right)
  \sum_{\I_n, \Ib_\nb} (n+1)(\nb+1)g^{i \ib} g^{j \jb} W_{ij\I_n}
  \Wb_{\ib \jb \Ib_\nb} \P^{i_1} \cdots \P^{i_n} \P^{\ib_1} \cdots
  \P^{\ib_\nb},
\end{split}
\end{equation}
where $g^{i \jb}$ denotes the uncorrected inverse K\"{a}hler metric at
the point $p$, the origin in field space.
For $n=\nb=1$, we get a one loop correction to the K\"{a}hler metric
at the origin $p$ arising from the dimensionless superpotential
couplings $A_{ijk}$,
\begin{equation}
  \label{eq:gcorr}
  \Delta g^{(1)}_{i\jb} \sim \left(\frac{1}{2\pi^2}
  \log{\frac{M_c^2}{M_{c'}^2}} \right) g^{k \kb} g^{l \lb} A_{ikl}
  \Ab_{\jb \kb \lb}. 
\end{equation}
Radiative stability requires that $\Delta g^{(1)}_{i \jb}$ is
``small'' compared to $g_{i \jb}$, that the coordinate-invariant norms
of tangent vectors to $X$ at $p$ under $\Delta g^{(1)}_{i \jb}$ are
smaller that those under $g_{i \jb}$, or roughly,
\begin{equation}
  \label{eq:glessthan}
  g^{j \jb} \Delta g^{(1)}_{i\jb} \sim \left(\frac{1}{2\pi^2}
  \log{\frac{M_c^2}{M_{c'}^2}}
  \right) g^{j \jb} g^{k \kb} g^{l \lb} A_{ikl} \Ab_{\jb \kb \lb}
  \lesssim  g^{j \jb} g_{i \jb} = \delta^j_i.
\end{equation}
We can re-phrase this requirement in precise, coordinate invariant
terms by taking traces and determinants of both sides,
\begin{equation}
  \label{eq:KboundsW}
  \det{\left[g^{j \jb} \Delta g^{(1)}_{i\jb} \right]}
\lesssim 1, ~~~~ \Tr{\left[ g^{j \jb} \Delta g^{(1)}_{i\jb} \right]}
\lesssim N.
\end{equation}
As promised, this gives us parametric bounds on the growth of the
dimensionless superpotential couplings $A_{ijk}$,
\begin{align}
  \label{eq:AIIIbounds}
 g^{i \ib} g^{j \jb} g^{k \kb} A_{ijk} \Ab_{\ib \jb \kb}
& \sim \mcO(N), \\
  \det{}_{ij}{\left[ g^{j \jb} g^{k \kb} g^{l \lb} A_{ikl} \Ab_{\jb \kb
  \lb}\right]} & \sim \mcO(1).
\end{align}
Since $A_{ijk}$ is a tensor, the interpretation of these bounds on the
size of its components is, of course, a coordinate dependent
question. Now, in a K\"{a}hler manifold, $d\omega = 0$ implies that it
is always possible (see \cite{GrifHar} p.107) to choose local
holomorphic coordinates about any smooth point $p$ such that
$\P^i(p)=0$ and
\begin{equation}
\del_i \delb_\jb K (\P, \Pb) = g_{i \jb}(\P,\Pb) = \delta_{i \jb} +
\mbox{terms of order $\ge 2$ in the $\P, \Pb$}.
\end{equation}
In these coordinates the above bounds simplify to,
\begin{equation}
  \label{eq:AIIIsimp}
  \sum_{i,j,k} |A_{ijk}|^2 \sim \mcO(N), ~~~~ 
\det{}_{ij}{\left[\sum_{k,l} A_{ikl} \Ab_{jkl} \right]} \sim \mcO(1).
\end{equation}
Thus, in these special coordinates, we see that in order to satisfy
these bounds, the generic components of $A_{ijk}$ must have
parametrically suppressed magnitudes at large $N$,
\begin{equation}
|A_{ijk}| \sim \mcO(N^{-1})
\end{equation}
though $\mcO(N)$ of them can still be as large as $\mcO(1)$ in
that limit. Therefore, we will refer to a tensor $A_{ijk}$ obeying
(\ref{eq:AIIIbounds}) as being {\em generically small}. 

We can similarly obtain bounds on the coefficients $A_{\I_n}$ and
$K_{\I_n \Ib_\nb}$ by considering the leading loop correction to the
quadratic term in the K\"{a}hler potential coming from the
corresponding higher order terms. One finds a slew of conditions
similar in form to \eqref{eq:AIIIsimp}. For instance,
\begin{subequations}\label{eq:KboundsII}
\begin{align}
  M_c^{2n-2} g^{i_1 \ib_1} \cdots g^{i_{n+2} \ib_{n+2}} A_{i_1 \cdots
    i_{n+2}} \Ab_{\ib_1 \cdots \ib_{n+2}} & \sim \mcO(N), \\
  \det{}_{ij}{\left[M_c^{2n-2} g^{j \jb } g^{i_1 \ib_1} \cdots
      g^{i_{n+1} \ib_{n+1}} A_{i_1 \cdots i_{n+1} i} \Ab_{\ib_1 \cdots
        \ib_{n+1} \jb} \right]} & \sim  \mcO(1), \\
  g^{i_1 \ib_1} \cdots g^{i_{n} \ib_{n}} K_{i_1 \cdots i_{n} \ib_1
    \cdots \ib_{n}} & \sim \mcO(N), \\
  \det{}_{ij}{\left[g^{j \jb } g^{i_1 \ib_1} \cdots g^{i_{n-1}
        \ib_{n-1}} K_{i_1 \cdots i_{n-1} i \ib_1 \cdots \ib_{n-1} \jb}
    \right]} & \sim \mcO(1).
\end{align}
\end{subequations}
In particular, this implies that $\mcO(N)$ of the coefficients
$A_{\I_n}$ could be $\mcO(1)$, while the generic coefficient must be
small,
\begin{equation}
  \label{eq:Anbounds}
|A_{i_1 \cdots i_n}| \sim M_c^{-(n-3)} \mcO(N^{-(n-1)/2}). ~~~~~ (n \ge 3)
\end{equation}
Thus, just as with $A_{ijk}$, we will refer to any tensor obeying bounds of
the form \eqref{eq:KboundsII} as {\em generically small}.  In general,
these conditions, as well as a very large number of others
corresponding to loop diagrams involving multiple vertices, constrain
all the coefficients of the higher order terms in the K\"{a}hler
potential and superpotential to be generically small at large
$N$.\footnote{We note that there is an additional constraint on the
  size of the coefficients coming from the requirement that the power
  series we have been writing down actually {\em absolutely converge}
  in a region of size $\sim M_c$. These constraints, while stronger
  than those coming from radiative stability,  only constrain the
  {\it asymptotics} of the $A_{\I_n}$ for large $n$, rather than
  constraining the terms at some particular order in $n$.}

\subsection{SUSY Vacua of the Renormalizable Wess-Zumino Model
  at Large $N$} \label{subsec:VacStruct}

We now wish to study the vacuum structure of this effective field
theory. As discussed in the previous subsection, this amounts to
studying the critical points of the superpotential, $\partial_{i}W=0$.
For large $N$, we expect that the number of critical points of $W$
within a polydisk of radius $M_{c}$ to scale exponentially with $N$.

Since we assume that the power series expansion for $W$ is absolutely
convergent in the polydisk, a field theorist might reasonably take the
approach of {\it truncating} the power series at some finite order and
looking for the critical points of the resulting polynomial. Of
course, we cannot hope that a truncation to any finite order
polynomial can be an accurate guide to the critical points in the
entire polydisk of radius $M_{c}$. At best, we will hope to obtain the
critical points inside some much smaller polydisk, of radius $M_{r}\ll
M_{c}$. Generically, however, we would not expect to find {\it any}
critical points inside this smaller polydisk. The criterion for
finding (trustworthy) critical points within this smaller polydisk for
small $N$ is well-known: we require that the coefficients of the linear
and quadratic terms in \eqref{eq:Wexp} be $A_{\I_1}\sim\CO(M_{r}^{2})$
and $A_{\I_2}\sim\CO(M_{r})$, respectively, where
\begin{equation}
 \frac{M_{r}}{M_{c}} = \epsilon \ll 1
\end{equation}
As long as the quartic and higher terms in the superpotential are not
anomalously large, they give negligible corrections to the critical
points determined by truncating $W$ to cubic order\footnote{As long as
  we are away from the discriminant locus to be discussed below}.

At large $N$, this is not quite sufficient. Even assuming that the
higher $A_{\I_n}$ are {\it generically small}, in the sense of the
previous subsection, we still need to require
\begin{equation}\label{eq:epsilonbound}
  \epsilon < \frac{1}{\sqrt{N}}
\end{equation}
in order for these higher-order corrections to be negligible. To see
this, we can look at the invariant quantity,
\begin{equation}
  \label{eq:dWbound}
(M_{c}^{-4}g^{i\overline{\jmath}}\partial_{i}W
\partial_{\overline{\jmath}}\overline{W})|_{\P=\P_{*}}, 
\end{equation}
where $\P_{*}$ is the critical point derived from the cubic
approximation of $W$. Using the generic smallness of the
$|A_{\I_n}|\sim M_{c}^{-(n-3)} N^{-(n-1)/2}$ and $|\P_{i*}|\sim M_{r}
= \epsilon M_{c}$, we see that the corrections grow parametrically
with $N$, unless \eqref{eq:epsilonbound} is satisfied.
In what follows, we will keep $\epsilon$ as a free parameter,
cognizant of the fact that it must be sufficiently small for the story
to work.

In non-supersymmetric theories, ensuring that under radiative
corrections the coefficients of the super-renormalizable terms in the
potential {\it remain }much smaller than the cutoff is called the {\it
  hierarchy problem}. In a supersymmetric theory, there are no
perturbative corrections to the superpotential. Nonetheless, the fact
that the coefficients of the super-renormalizable terms in the
superpotential are of order $M_{r}$, rather than $M_{c}$, means that
the point about which we are expanding is, in some sense, ``special.''
That will be further brought home in \S\ref{subsec:R}, where we will
assume that the point $\Phi=0$ will be a point where there is an
unbroken $\BZ_{4}$ R-symmetry.

\subsection{Distributions of Models and Flux Compactifications of IIB}
\label{subsec:Flux}

Since F-theory compactifications with flux are the prime motivating
example of ``landscape'' models with a large number of vacua, let us
pause to consider how such models fit in with our general
considerations, as developed so far. Any {\it given} set of fluxes
satisfying the requisite tadpole cancellation conditions in an
F-theory or IIB Orientifold compactification gives rise to a
Gukov-Vafa-Witten \cite{Gukov:1999ya} superpotential for the complex
structure moduli.  Since there are, in general, many solutions to the
tadpole cancellation conditions, we have {\it an ensemble of theories}
with different superpotentials, labeled by the possible fluxes.

In the IIB case, the elements of the ensemble are
labeled by all choices of integer fluxes through the $2b^{2,1} + 2$
three cycles $\Sigma_a$ of the Calabi-Yau 3-fold $\mcM$,
\begin{equation}
  \label{eq:Fluxes}
\frac{1}{(2 \pi)^2 \alpha'} \int_{\Sigma_a} F = N_a \in \Z, ~~~~
\frac{1}{(2 \pi)^2 \alpha'} \int_{\Sigma_a} H = M_a \in \Z.
\end{equation}
compatible with the tadpole cancellation condition for induced
D3-brane charge on $\mcM$, 
\begin{equation}
  \label{eq:tadpole}
  Q_3 = \frac{1}{(2\pi)^4 \alpha'^2}\int_\mcM F \wedge H.
\end{equation}
where $Q_3$ includes the contribution of space-filling, mobile
D3-branes, D7-branes, and O3-planes. Now, if $Q_3 \sim b_{2,1} \sim
\mcO(N)$ then tadpole cancellation requires that each of the integer
fluxes can then be at most $N_a \sim M_a \sim \mcO(N^{1/2})$. Each
choice of fluxes gives rise to a Gukov-Vafa-Witten superpotential for
the IIB axio-dilaton $\tau$ and the complex structure moduli $z_i =
\int_{A_i} \Omega$ of $\mcM$,
\begin{equation}
  \label{eq:GVW}
  W = \int_\mcM G \wedge \Omega.
\end{equation}
where $G=F-\tau H$ and the $A_i$ and $B^i$ form a symplectic basis of
3-cycles $\{ \Sigma_a \} = \{ A_i, B^i \}$ in $\mcM$ with respect to
the intersection pairing on $H_3(\mcM, \Z)$. In particular, we see
that the tadpole cancellation condition places bounds on the growth of
$W$ on $N$, 
\begin{equation}
  \label{eq:GVWbound}
  W \supset \sum_i \int_{B^i} G \int_{A_i} \Omega \sim \sum_i (N^i - \tau
  M^i) z_i \sim \mcO(N).
\end{equation}
which is certainly consistent with the condition that the couplings of
the moduli are indeed generically small. However, it is not at all
clear that there is any point in the moduli space about which the
cubic truncation gives a good approximation to the locations of its
critical points -- that is, we do not expect the renormalizable
approximation to be a useful guide to the vacuum structure.

Further, notice that the superpotential is odd under $G \rightarrow -
G$, while the tadpole cancellation condition is even, so if $G$
satisfies the tadpole constraint, then so does $-G$. This means that
if we find a supersymmetric vacuum $z_i^*$, a critical point of
(\ref{eq:GVW}) with a given flux $G$, and a corresponding value $W^*$
of the superpotential, then $z_i^*$ also corresponds to a vacuum of
the model with flux $-G$ and superpotential $-W^*$.  Thus, even though
the average value of $W$ at the supersymmetric vacua for any fixed
choice of the fluxes may not vanish, the {\em ensemble} average of $W$
certainly does vanish.

Motivated by this example, we will be interested in situations where
we are actually given a distribution or ensemble of models. That is,
we will assume that physics at high energies $\sim M_p$ can be
understood as providing a distribution of coefficients $A_{I_n}$ for
the effective low-energy $\lesssim M_c$ models described above.
Indeed, in the IIB flux vacua, the physics which determines the values
of the quantized fluxes is not even field theoretic in nature - it
likely involves high energy string/brane dynamics, topology change,
etc. In particular, the scanning of the cosmological constant in these
models is explained by high energy physics. Certainly, we could accept
such a high energy explanation and restrict our consideration to
distributions of coefficients which are symmetric under $A_{I_n}
\rightarrow -A_{I_n}$ and share this property of flux vacua.  However,
we will focus on situations where there may be a low energy
explanation for this scanning.\footnote{Some high energy input,
  however, may be inevitable.  For example, one might worry about the
  origin of the small parameter $\epsilon = M_r/M_c$ that we required
  in order to make the cubic approximation. This appears to be a
  tuning of $\mcO(N)$ relevant couplings in the model. Without some
  further explanation, one might worry that, together, these
  represents a fine-tuning of order $\epsilon^N$, which would wipe out
  whatever ``advantage'' we gained in having $2^{N}$ vacua. This need
  not be the case if the smallness of each of the $\mcO(N)$ couplings
  has a common explanation. Perhaps there is an approximate symmetry,
  broken weakly by the effects that generate the superpotential, which
  guarantees that these couplings are small. Indeed, such a symmetry
  would be reflected in symmetries of the resulting probability
  distribution for the couplings. For example, if the coefficients of
  the relevant operators were all selected from the {\em same}
  distribution, we would indeed only require a single fine tuning of
  the distribution. }  One way to achieve this
\cite{Arkani-Hamed:2005yv} is through the imposition of an R-symmetry,
which we discuss presently.

\subsection{R-Symmetry, The Renormalizable Wess-Zumino Model, and
  $\Lambda$}\label{subsec:R}

The general renormalizable Wess-Zumino model of $N$ interacting chiral
superfields $\P^i$ is described by a quadratic K\"{a}hler potential,
\begin{equation}
  \label{eq:KpotWZ}
K(\P, \Pb) = g_{i \jb} \P^i \Pb^\jb,
\end{equation}
and a cubic superpotential,
\begin{equation}
  \label{eq:superpot}
W(\P^i) = A + A_i \P^i + \frac{1}{2} A_{ij} \P^i \P^j + \frac{1}{3}
A_{ijk} \P^i \P^j \P^k.
\end{equation}
Radiative stability of the scaling of the K\"{a}hler potential at
large $N$ restrict us to the case that generically $|A_{ijk}| \sim
N^{-1}$ with $\mcO(N)$ terms which are $\mcO(1)$. The supersymmetric
vacua for this theory are points where,
\begin{equation}
  \label{eq:SUSYvacWZ}
\del_i W = A_i + A_{ij} \P^j + A_{ijk} \P^j \P^k=0,
\end{equation}
a set of $N$ {\em complex algebraic} equations in the $\P^i$. In
particular, these polynomials determine an ideal\footnote{ See section
  \ref{sec:Solve} for a basic review of the commutative algebra
  language used here, and \cite{CLO, CLO2} for a more detailed
  introduction.} in the ring of polynomials in the $\P^i$, $\< \del_i
W \> \subset \C[\P^1, \ldots, \P^N ]$. Bezout's theorem (see Chapter
3, Theorem 5.5 of \cite{CLO}) guarantees that for a \emph{generic}
choice of the coefficients of $W$,\footnote{By generic, we mean that a
  certain polynomial in the coefficients known as a resultant is
  non-vanishing -- see \S\ref{subsubsec:generic}.}  these $N$ simultaneous
quadratic equations have $2^N$ roots.  From an algebraic point of
view, this translates into the fact that the quotient ring,
\begin{equation}
  \label{eq:quotring}
  \C[\P^1, \ldots, \P^N] / \< \del_i W \> \cong \C^{2^N}
\end{equation}
is a $2^N$-dimensional vector space over $\C$ (see Chapter 3, Theorem
6.2 of \cite{CLO}) generated by the images of the monomials $\P^{i_1}
\cdots \P^{i_r}$ with $i_1 < \cdots < i_r \le N$. Explicitly, the
quotient ring should be understood as the ``polynomial functions'' on
the algebraic variety cut out by the equations generating the ideal,
which in this case are the functions on a set of $2^N$ points, i.e. a
$2^N$ dimensional vector space.

Further, we saw that the critical points are generically contained in
a neighborhood $U$ of radius $\sim M_{r}$ about the origin. Note that if
we assume that the constant term $A$ in $W$ is also of order $\lesssim
N M_{r}^3$, the superpotential at each of the critical points in $U$ is
roughly of the order of\footnote{As a test, we note that the
  prototypical example of a landscape superpotential, the
  Gukov-Vafa-Witten superpotential for the complex structure moduli of
  a IIB orientifold model indeed scales this way - see
  (\ref{eq:GVWbound}).  Again, in a appropriate basis, the nonzero
  coefficients in the Gukov-Vafa-Witten superpotential can be $\sim
  \CO(1)$, but, they are, indeed, very sparse.}
\begin{equation}
  \label{eq:sizeW}
W \sim M_{r}^3 \times\mcO(N).
\end{equation}
This gives a supersymmetric contribution to the vacuum energy
\begin{equation}
  \label{eq:vacenergy}
   \Lambda = - 3 \frac{|W|^2}{M_p^2}\left(1 + \mcO(M_{r}^2/M_p^2)
   \right)  \sim - N \epsilon^{2} M_r^4,
\end{equation}
where $\epsilon = M_{r}/M_{c}\lesssim 1/\sqrt{N}$ and
$M_{c}/M_{p}\lesssim 1/\sqrt{N}$.  Note that all corrections coming
from the inclusion of the Chern Connection terms and higher order
terms in the K\"{a}hler potential and superpotential are
parametrically suppressed at large $N$. 

When supersymmetry is broken, we get a positive contribution,
$\Lambda_{S}$ to the vacuum energy. The hope is that the distribution
of values for $W$ among the $2^{N}$ vacua will be such that
\eqref{eq:vacenergy} can very nearly cancel
$\Lambda_{S}$.\footnote{See \cite{ Douglas:2004qg,
    Arkani-Hamed:2004fb, Arkani-Hamed:2004yi, Susskind:2004uv,
    Kallosh:2004yh, Dine:2004is, Dine:2004ct, Dine:2005iw} for
  discussions regarding the scale of supersymmetry breaking in
  landscape models.} To achieve this \cite{Arkani-Hamed:2005yv}, one
wants a distribution of values of $W$, such that the standard
deviation is large, compared to the mean value, $\vvev{W}$.

As the authors of \cite{Arkani-Hamed:2005yv} noted, this is easy to
arrange. If the superpotential is odd under $\P^i \rightarrow - \P^i$,
then supersymmetric vacua come in pairs ${\P^i}^*$ and $-{\P^i}^*$
with opposite values of $W$ and we would have $\vvev{W}=0$. We can
enforce this by imposing a $\Z_4$ R-symmetry $\P^i(y,\theta)
\rightarrow -\P^i(y,i\theta)$, under which the superpotential must
have charge 2 and therefore is an odd polynomial,
\begin{equation}
  \label{eq:superpotR}
W(\P^i) = \sum_{i=1}^N A_i \P^i + \frac{1}{3} \sum_{i,j,k=1}^N A_{ijk}
\P^i \P^j \P^k.
\end{equation}
Such an R-symmetry is clearly non-generic. In particular, the origin
in these coordinates must be a special point for such a symmetry to
hold, as an expansion of the superpotential at any nearby point in
field space certainly would not exhibit the same R-symmetry.  That is,
the R-symmetry is spontaneously broken by a vev for the $\P^i$.  In
fact, given an arbitrary SUSY non-linear sigma model, there is no
reason to believe that its superpotential will generically ever have a
point in the target space where $W$ has such a symmetry.  Thus, the
imposition of an R-symmetry means that we are restricting our
consideration to very special SUSY non-linear sigma models expanded
locally about a special point in their target spaces. This is the
price that one must pay for a low-energy explanation for the scanning
of the vacuum energy. For further discussion regarding this issue, see
\cite{DeWolfe:2004ns, Dine:2004dk, Dine:2005gz}. Of course, R-symmetry
is phenomenologically desirable for many other reasons as well (see
the discussion in \cite{Arkani-Hamed:2005yv}).

The algebraic consequences of the R-symmetry and their geometric
interpretations will be of use in our discussion of the statistics of
the vacua of this model.  Note that the conditions for unbroken SUSY
are actually equations for the scalars $\p^i$ in the chiral
superfields $\P^i$ and the R-symmetry acts as a $\Z_2$ parity on these
scalars, $\p^i \rightarrow -\p^i$. In particular, the SUSY vacua of
this model are determined by $N$ quadratic equations,
\begin{equation}
  \label{eq:SUSYvacR}
\del_i W = A_i + \sum_{j,k=1}^N A_{ijk} \p^j \p^k = 0
\end{equation}
which are invariant under the $\Z_2$ parity. As a result, these
quadratics actually determine an ideal in the ring of $\Z_2$
invariants constructed from polynomials in the $\p^i$,
\begin{equation}
\<\del_i W\> = \< A_i + A_{ijk} \p^j \p^k \> \in \C [\p^1, \ldots,
\p^N]^{\Z_2}.
\end{equation}
We can interpret the ring of invariants $\C[\p^1, \ldots, \p^N
]^{\Z_2}$ as the ``polynomial functions'' on the orbifold $\C^N/\Z_2$.
Then, in analogy with the general case, we can consider the quotient
ring,
\begin{equation}
  \label{eq:quotringR}
  \C[\p^1, \ldots, \p^N]^{\Z_2} / \< \del_i W \> \cong \C^{2^{N-1}},
\end{equation}
which is a $2^{N-1}$-dimensional vector space over $\C$ generated by
the images of the even monomials $\P^{i_1} \cdots \P^{i_{2r}}$ with
$i_1 < \cdots < i_{2r} \le N$. Geometrically, this quotient ring can
be interpreted as the polynomial functions on the variety cut out by
the $\del_i W$ in the orbifold, which consists of the $2^{N-1}$ images
of the $2^{N}$ critical points of $W$ in $\C^N/\Z_2$. 

\subsection{Fermat Form of the Cubic}\label{subsec:Fermat}

We are interested in the properties of distributions of models with
superpotentials respecting the R-symmetry (\ref{eq:superpotR}) and
quadratic K\"{a}hler potentials (\ref{eq:KpotWZ}). At first glance, one
might expect to describe such a distribution of models as an arbitrary
probability distribution on the space of all superpotential and
K\"{a}hler couplings, $f(A_i,A_{ijk},g_{i \jb})$. However, we should
not distinguish between models which differ by field
redefinitions. Of course, arbitrary field redefinitions will
not preserve the form of the Wess-Zumino models we are considering.
However, it is easy to see that the field redefinitions which do
are of the form $\p^i \rightarrow G^i_m \p^m$, where $G^i_m \in GL_N
(\C)$. In particular, they act on the space of couplings as,
\begin{equation}
  \label{eq:GLNac}
A_i \rightarrow A_m G^m_i, ~~~~ A_{ijk} \rightarrow A_{mnp} G^m_i
G^n_j G^p_k, ~~~~ g_{i \jb} \rightarrow g_{m \nb} G^m_i
\bar{G}^\nb_\jb.
\end{equation}
If we posit some probability distribution on the space of {\it
  couplings}, this distribution gets averaged over $GL_{N}(\C)$ orbits
to produce a probability distribution on the space of {\it theories}.
That is awkward to deal with. If possible, it is much better to fix
the redundancy by choosing a gauge for the $GL_N(\C)$ field
redefinitions. Since $GL_N(\C)$ is $N^2$ dimensional, such a gauge
condition should involve precisely $N^2$ independent, complex
algebraic conditions.  Now, in working with algebraic equations, it is
often convenient to redefine our variables in order to make the
coefficients of the terms of highest degree as simple as possible. If
$W$ is a generic polynomial of degree $d$, we can use the $GL_N(\C)$
symmetry to enforce $N^2$ conditions on $A_{\I_d}$,
\begin{equation}
  \label{eq:GaugeCond}
A_{i \cdots i} = 1, ~~~~ A_{ij \cdots j} = A_{jij \cdots j} = \cdots =
A_{j \cdots j i} = 0,~~ \mbox{for $i \neq j$}.
\end{equation}
We will refer to a polynomial satisfying these conditions as being in
Fermat form, 
\begin{equation}
  \label{eq:FermatForm}
  W = \frac{1}{d} \left( (\p^1)^d + \cdots + (\p^N)^d \right) +
  (\mbox{terms of degree $\le d-1$ in each $\p^i$}).
\end{equation}
Note that if $W$ is {\em odd} (so it has a $\Z_2$ R-symmetry as
above), then the Fermat form is somewhat stronger, as the additional
terms in (\ref{eq:FermatForm}) actually have degree at most $(d-2)$ in
any $\p^i$. In particular, for the case of interest here, $d=3$, with
the R-symmetry above, the superpotential takes the form,
\begin{equation}
  \label{eq:FermatCubic}
  W(\p^i) = \sum_{i=1}^N \left( \tfrac{1}{3} (\p^i)^3 - a_i \p^i \right) 
- \sum_{i < j < k}^N b_{ijk} \p^i \p^j \p^k,
\end{equation}
where $a_i$ has mass dimension 2 and is generically $M_{r}^2\times
\mcO(1)$ and the $b_{ijk}$ are symmetric, traceless (so $b_{iii} =
b_{ijj} = b_{jij} = b_{jji}=0$), dimensionless and generically
$|b_{ijk}| \sim \mcO(N^{-1})$ with at most $\mcO(N)$ of them
$\mcO(1)$. In going to Fermat form, it is important to note that we
cannot also simultaneously set $g_{i \jb} = \delta_{i \jb}$. Further,
the genericity of $W$ is important here --- not all polynomials can be
put into Fermat form. For example, the following model with
spontaneous breaking of supersymmetry via the O'Raifeartaigh
Mechanism,
\begin{equation}
  \label{eq:ORaif}
  W = Z (X^2 - a) + Y X^2
\end{equation}
cannot be put into this form. In fact, it is easy to show by analytic
computations using eigenvalue methods (see \S\ref{sec:Solve})
that for $N=3$ and $d=3$ no polynomial in Fermat form exhibits the
O'Raifeartaigh Mechanism.

We believe that this may be true much more generally. More precisely,
we conjecture that {\em all} cubic, R-symmetric superpotentials of the
Fermat form (\ref{eq:FermatCubic}) have at least a pair of
supersymmetric vacua. While a proof of this fact is far beyond the
scope of this work, we hope to return to this question in the future.

\subsection{Symmetries of the Fermat Form}\label{subsec:symmetries}

Just as is often the case in gauge fixing, it is important to note
that our gauge condition (\ref{eq:GaugeCond}) does not completely fix
the $GL_N(\C)$ redundancy. For any generic $A_{\I_d}$, there is a finite
subgroup $H$ of $GL_N(\C)$ transformations which transforms the
coefficients in such a way as to keep $A_{\I_d}$ in the Fermat form
(\ref{eq:FermatForm}).  Namely, $h^m_i \in H$ are the solutions to
$N^2$ algebraic equations of degree $d$ in the $N^2$ complex
components $h_m^i$,
\begin{equation}
  \label{eq:Symmetry}
A_{m_1 \cdots m_d} h^{m_1}_{i} \cdots h^{m_d}_{i} = 1, ~~~~ A_{m_1
  \cdots m_d} h^{m_1}_{i} h^{m_2}_{j}  \cdots h^{m_d}_{j} = 0, ~~
  \mbox{for $i \neq j$}.
\end{equation}
If these were generic degree $d$ equations, Bezout's theorem would
tell us that they have $d^{N^2}$ solutions.  However, it turns out
that this is not the case for generic $A_{\I_d}$, and $d^{N^2}$ is
actually a strict upper bound on the order of $H$. We will show this
by exhibiting a large subgroup of $H$ which is independent of
$A_{\I_d}$ and whose order does not divide $d^{N^2}$. First, note that
permutations of the $\p^i$ certainly won't take us out of Fermat
form, so we expect that the symmetric group on $N$ variables $S_N
\subset H$. Next, note that multiplying any $\p^i$ by a $d^{th}$ root
of unity also won't ruin the special form, so we get a subgroup of $H$
isomorphic to $(\Z_d)^N$ given by,
\begin{equation}
  \label{eq:zdsymm}
  h_i^j = \delta_i^j \zeta^{n_i}, ~~~~ \zeta = e^{2\pi i/d}, ~~~~ 1 \le
  n_i \le d. 
\end{equation}
Thus, we see that $S_N \ltimes (\Z_d)^N$, which has order $N! \times
d^N$, must be a subgroup of H. Since $N!$ generally does not divide
$d^{N^2}$, we see that,
\begin{equation} 
N! \times d^N \le |H| < d^{N^2}
\end{equation}
In the simplest non-trivial case, $N=3$ and $d=3$, we numerically
solved (\ref{eq:Symmetry}) using Mathematica for various values of the
coefficients and found that $H$ is generically of order $648 = 2^3
\times 3^4 = 3! \times 3^3 \times 4$. In particular, there are indeed
elements of $H$ which depend on $A_{\I_d}$, which suggests that $|H|$
may generally be strictly larger than $S_N \ltimes (\Z_d)^N$. This is
not unexpected. Rather generally, the space of common solutions of
these polynomial equations is an algebraic variety. Different values
for the coefficients of the polynomials will nonetheless lead to
isomorphic algebraic varieties. $H$ is the modular group, and we will
not attempt to give a full characterization of $H$ beyond its
``obvious''$S_N\ltimes (\Z_d)^N$ subgroup. We will content ourselves
with exploiting the constraints that stem from this latter subgroup.

To summarize, after fixing gauge and imposing R-symmetry, each model
is uniquely described up to a discrete symmetry $H$ containing $S_N
\ltimes (\Z_3)^N$ by a cubic superpotential in Fermat form
(\ref{eq:FermatCubic}) and a quadratic K\"{a}hler potential
(\ref{eq:KpotWZ}). Thus, a distribution of field theory landscape
sectors is a probability distribution on the space of couplings,
\begin{equation}
  \label{eq:InvtDist}
f(a_i, b_{ijk}, g_{i \jb}) d a_i d b_{ijk} d g_{i \jb}
\end{equation}
which is invariant under $H \supset S_N \ltimes (\Z_3)^N$, 
\begin{equation}
  \label{eq:Hact}
  a_i \rightarrow a_m h^m_i, ~~~~~ b_{ijk} \rightarrow b_{mnp} h^m_i
  h^n_j h^p_k, ~~~~~ g_{i \jb} \rightarrow g_{m \nb} h^m_i \bar{h}^\nb_\jb.
\end{equation} 
We assume that this distribution is fixed by high energy
physics. Lacking any clear physical input which distinguishes
among the different vacua of a {\em given} low-energy model, we
simply treat these vacua democratically. In the following sections, we
use algebraic methods to compute some statistical properties of
various physical quantities averaged over the $2^N$ vacua of each
model as a function of the couplings. Of course, these
model averages can then be further averaged over the ensemble using
the high energy distribution function. 

\section{Eigenvalue Methods for Solving Algebraic Equations}\label{sec:Solve}

The solution of $N$ simultaneous linear equations in $N$ variables is
something we all learn to do from an early age using several different
methods. For example, we may decide to do it via Gaussian elimination
or via matrix methods - using determinants and Cramer's rule to invert
the matrix of coefficients. A natural question to ask is if there
exist analogues of these methods which could be used to solve
simultaneous equations of higher degree. Indeed, this is the case. The
method of Gaussian elimination has a vast generalization in the study
of algebraic elimination theory and Gr\"{o}bner bases methods (see
\cite{CLO, CLO2}). These methods provide algorithms which allow one to
systematically eliminate variables one by one in any given system of
equations at the cost of increasing the degrees of the resulting
system of equations in fewer variables and, with some further work,
find (approximate) solutions. Unfortunately, these are not directly
useful for us as we will be interested in the statistical properties
(such as the average and variance of the superpotential and other
observables) of such solutions as a function of the coefficients -
i.e. in the properties of solutions of {\em families} of such
equations.  It turns out that the generalization of matrix methods to
the case of simultaneous equations of higher degree does end up being
quite useful for these purposes.

\subsection{Rings, Ideals and a Toy Model - Quadratic $W$}

We will introduce these methods by applying them to the simple
case of a {\em generic} quadratic superpotential,
\begin{equation}
  \label{eq:quad}
  W = W_0 - A_i z_i + \frac{1}{2} C_{ij} z_i z_j = W_0 - A \cdot z +
  \frac{1}{2} z \cdot C \cdot z.
\end{equation}
Of course, the ``statistics of vacua'' in this case may seem to be,
well, vacuous - the critical point of this polynomial is given by the
unique solution of $N$ linear equations,
\begin{equation}
  \label{eq:critquad}
\del_i W = -A_i + C_{ij} z_j = 0 \Rightarrow C \cdot z = A,
\end{equation}
which is, of course,
\begin{equation}
  \label{eq:linsol}
  z = C^{-1} \cdot A,
\end{equation}
where the genericity condition is that $\det C \neq 0$, or $C \in GL_N(\C)$.
However, if we are given a {\em distribution} of coefficients for $W$,
we might be interested in the corresponding distribution of the values
of $W$ at the critical point. To attack such problems more generally,
it is essential (though, in this case, akin to trying to kill an ant
with a machine gun) to recast these questions in the algebraic
language of polynomial rings and ideals.

To begin with, let us quickly review what we will need about ideals,
rings, and varieties. Roughly, a ring is a set $R$ whose algebraic
properties mimic many of those of the integers $\Z$. That is, we may
add, subtract, and multiply elements of $R$ and obtain new elements in
$R$, though the same may not be true for division. Recall that an
ideal $I$ contained in a ring $R$ is defined by the property that if
$f, g \in I$ and $r, s \in R$ then $rf + sg \in I$.  In particular,
this implies that the {\em quotient ring} of $R$ by $I$, $Q=R/I$ where
we identify all elements of $R$ differing by an element in $I$, is
well-defined. Note that all the rings that we consider are {\em
Noetherian} rings, which means that all their ideals are finitely
generated, so
\begin{equation}
  \label{eq:ideal}
I = \< f_1, \ldots, f_m \> = \left\{ \left. \sum r_i f_i \right| r_i
\in R \right\} = \< f_i \>.
\end{equation}
The rings we will be most interested are rings of polynomial functions
in $N$ variables, $R = \C[z_1, \ldots, z_N]$, and their quotients by
ideals $I$ generated by some polynomials $f_i(z_1, \ldots, z_N) \in
R$. Thus, $I$ is just the set of all polynomials which vanish on the
zero locus in $\C^N$ of the simultaneous set of polynomial equations
$f_i(z_j) = 0$, $i = 1, \ldots, m$. This zero-locus is called the {\em
algebraic variety} in $\C^N$ corresponding to the ideal $I$. Thus,
non-vanishing elements of the quotient ring $Q=R/I$ are precisely the
non-trivial residues of polynomial functions on $\C^N$ restricted to
the variety - they define a notion of ``polynomial functions'' on the
variety.

As we have already mentioned above, the polynomials $\del_i W$ which
determine the critical point generate an ideal in the ring of
polynomials in $N$ variables,
\begin{equation}
  \label{eq:idealWquad}
  I = \< \del_i W \> = \< -A_i + C_{ij} z_j \> = \< -A + C \cdot z\>.
\end{equation}
Clearly, any linear combination of the generators $\del_i W$ with
complex coefficients is also in the ideal. In particular, if $\det C
\neq 0$ then $C$ is invertible and we can consider the $N$ elements of
$I$, $z - C^{-1} \cdot A$, obtained by multiplying the $N$ generators
$\del_i W$ by the matrix $C^{-1}$. In fact, as $C$ is invertible,
these are also generators of the ideal $I$. Thus, in the quotient
ring $Q=\C[z_1, \ldots, z_N] / I$, we can use the $N$ relations $z_i -
{C^{-1}}_{ij} A_j=0$ to systematically eliminate the $z_i$ and rewrite
any element of $Q$ in terms of multiples of the identity -
constants. So, we see that
\begin{equation}
  \label{eq:ringquad}
  Q = \C[z_1, \ldots, z_N] / I \cong \C,
\end{equation}
a one dimensional complex vector space, generated by constants. In
particular, the residue of any polynomial $f(z_1, \ldots, z_N) \in
\C[z_1, \ldots, z_N]$ in $Q$ is obtained by setting to zero all
elements of the ideal $I$, that is, by substituting $z = C^{-1} \cdot
A \in \C^N$ into $f$.  Thus, we see that {\em the residue of $f$ in
$Q$ is precisely its value at the critical point of $W$}, and that the
variety corresponding to $I$ is the single critical point $z = C^{-1}
\cdot A \in \C^N$. Indeed, as the only functions on a point are
constant functions - the value at the point - this is consistent with
our intuition that the quotient ring $Q=\C[z_1, \ldots, z_N] / I$
should be interpreted as the ``polynomial functions'' on a
point. Further, note that as $Q$ is isomorphic as a ring to $\C$, the
product structure of the ring of polynomials is preserved in the
quotient. That is, we can think of the residue of $f$ in $Q$ as an
operator on other elements of $Q$ which is multiplication by its value
at the critical point. In particular, using the fact that $\del W, z -
C^{-1} \cdot A \in I$, we can compute the residue of $W$ in $Q$ by,
\begin{equation}
W = W_0 - \frac{1}{2} A \cdot z + \frac{1}{2} z \cdot \del W \sim W_0
- \frac{1}{2} A \cdot z \sim W_0 - \frac{1}{2} A \cdot C^{-1} \cdot A
\in Q.
\end{equation}
Of course, we could have just plugged in our solution, but it turns
out that this kind of computation generalizes to the case of higher
degree in a way that is useful for computing statistical properties.

\subsubsection{Non-Generic Quadratic Superpotentials}

It is also interesting to ask what happens if $C$ is {\em not}
generic, if $\det C = 0$. Then, $C$ has a non-trivial null space which
is the space solutions of the homogeneous linear equations we get in
the limit that we take $A_i \rightarrow 0$. We can describe this limit
more precisely by lifting the equations to projective space, to
$\CP{N}$. That is, we add a new coordinate $z_0$ (which we may think
of physically as a chiral superfield corresponding to a (complexified)
scale factor) and then consider the homogeneous system of $N$ linear
equations $C_{ij} z_j - A_i z_0 = 0$ up to (complex) scaling in
$\CP{N}$.  In the open set $U_0 = \{z_i \sim \lambda z_i \in \C^{N+1}-
\{0\} | z_0 \neq 0 \} \subset \CP{N}$, the $u_i = \frac{z_i}{z_0}$ are
good coordinates and the above equation reduces to the inhomogeneous
equation with $z_i \rightarrow u_i$. Thus, if $\det C \neq 0$, the
unique solution of the inhomogeneous equation in $U_0$ gives us the
unique solution in $\CP{N}$ as well. However, the advantage of
projectivizing the problem is that we have added points corresponding
to solutions even in the non-generic case. If $\det C=0$ then we have
at least one solution in $\CP{N}$ which has $z_0 =0$ given by a
one-dimensional subspace of the null space of $C$. As the $u_i
\rightarrow \infty$ as we take $z_0 \rightarrow 0$, we can think of
such a solution as ``a solution at infinity'' from the perspective of
$U_0$. Thus, we see that non-generic points in the space of couplings
of the quadratic superpotential correspond to situations in which the
supersymmetric vacuum has ``run off to infinity'' in field space. Now,
since all of our field theory computations are approximations only
valid locally in field space, these non-generic points should more
properly be understood physically as points in the space of couplings
in which those approximations break down.

\subsection{The General Case}

Now, let's consider the generalization of the above to the case that
$W$ is of degree $d>2$,
\begin{equation}
  \label{eq:supWz}
W(z_1, \ldots, z_N) = \sum_{n=1, \I_n}^d \frac{A_{\I_n}}{n} z_{i_1}
\cdots z_{i_n}.
\end{equation}
The critical points of $W$ are then given by the simultaneous
solutions of $N$ degree $d-1$ polynomials, $\del_i W = 0$, and, as we
have already mentioned above, the polynomials $\del_i W$ generate an
ideal in the ring of polynomials in $N$ variables,
\begin{equation}
  \label{eq:idealWz}
  I = \< \del_i W \> = \< \sum_{n=1, \I_{n-1}}^d A_{i\I_{n-1}} z_{i_1} \cdots
  z_{i_{n-1}} \>.
\end{equation}
Just as before, we may now try to simplify the generators by taking
$\C$-linear combinations of them. In particular, it would be nice to
simplify the form of the terms of highest degree as we did in
obtaining the Fermat form. So, define the matrix $C$ by,
\begin{equation}
  \label{eq:CMatrix}
A_{i \cdots i} = C_{ii}, ~~~~ A_{ij \cdots j} = A_{jij \cdots j} = \cdots =
A_{j \cdots j i} = C_{ij},~~ \mbox{for $i \neq j$}.
\end{equation}
Now, just as in the case of the quadratic superpotential, if $\det C
\neq 0 \Rightarrow C \in GL_N(\C)$, we can multiply the $\del_i W$ by
the $C^{-1}$ and obtain a simpler set of generators,
\begin{equation}
  \label{eq:idealWzs}
  I = \< \sum_{n=1, \I_{n-1}}^d \sum_{j=1}^N
  C^{-1}_{ij} A_{j\I_{n-1}} z_{i_1} \cdots z_{i_{n-1}} \> = \< z_i^{d-1} +
   (\mbox{terms of degree $\le d-2$ in any $z_j$}) \>.
\end{equation}
Of course, if $W$ is in Fermat form, then $C_{ij} = \delta_{ij}$ and
so the generators $\del_i W$ themselves take precisely this
form.\footnote{Now, while it is obvious that a superpotential in
  Fermat form has $\det C \neq 0$, one might wonder if this condition
  is in general necessary and sufficient for the existence of a
  $GL_N(\C)$ coordinate transformation taking any given superpotential
  to Fermat form. However, this is not at all clear, as $C$ involves
  only nearly diagonal components of $A_{\I_d}$ while the
  transformation to Fermat form (\ref{eq:GLNac}) involves all the
  components of $A_{\I_d}$. We will not need to delve into this issue
  further for our purposes.} As we have argued earlier, the Fermat
Form is convenient for other reasons as well, so we will assume
henceforth that $W$ is given in Fermat form. Just as before, the above
form suggests that in the quotient ring $Q = \C[z_i, \ldots, z_N] / I$
we should be able to use the generators in (\ref{eq:idealWzs}) to
systematically eliminate all powers of $z_i$ greater than $(d-2)$ and
write any element of $Q$ only in terms of monomials with powers of
$z_i$ less than or equal to $(d-2)$. This is in fact the case as long
as $W$ is a {\em generic} polynomial of degree $d$ (for a rigorous
proof of this fact, see Chapter 3, Theorem 6.2 of \cite{CLO}). It is
not difficult to show this explicitly in the case of interest $d=3$,
but the result is not particularly enlightening and will not be
presented here. We only note here that the computation requires that
several large matrices in the coefficients be invertible, which we
believe is a computational manifestation of the assumption of
genericity. For the following, we will proceed assuming that $W$ is
generic, and discuss what we mean by this more precisely at the end of
the discussion.

Thus, for $W$ generic, the monomials $z_{1}^{d_1} \cdots z_{N}^{d_N}$
with $0 \le d_i \le d-2$ form a basis for $Q$ as a vector space, and
the relations will allow us to express the residue of any polynomial
$f(z_1, \ldots, z_N) \in \C[z_1, \ldots, z_N]$ in $Q$ in terms of this
basis. This is particularly simple in the case of the superpotential
itself, as we have
\begin{equation}
  \label{eq:superpotQ}
  W= \frac{1}{d} \sum_i z_i \del_i W + \sum_{n=0,\I_{n}}^{d-1}
  \frac{d-n}{d} A_{\I_n} z_{i_1} \cdots z_{i_n} \rightarrow
  \sum_{n=0,\I_{n}}^{d-1} \frac{d-n}{d} A_{\I_n} z_{i_1} \cdots z_{i_n} \in Q.
\end{equation}
Thus, we see that the quotient ring,
\begin{equation}
  \label{eq:quotringz}
  \C[z_1, \ldots, z_N] / \< \del_i W \> \cong \C^{(d-1)^N},
\end{equation}
is a $(d-1)^N$-dimensional vector space over $\C$ generated by the
images of the monomials $z_{1}^{d_1} \cdots z_{N}^{d_N}$ with $0 \le
d_i < d-1$. Further, as we mentioned earlier, this quotient ring
should be understood as the ``polynomial functions'' on the variety
cut out by the generators of the ideal, which in this case are the
functions on a set of $(d-1)^N$ points. In particular, in analogy with
with the quadratic case, we expect that the residue of any polynomial
$f(z_1, \ldots, z_N) \in \C[z_1, \ldots, z_N]$ in $Q$ should be
related to the values of $f$ at the critical points of $W$. However,
the fundamental difference between the case $d>2$ and the simple case
of a quadratic superpotential is that here,\ {\em the residue of $f$
  in $Q$ is a vector}, and it is not immediately obvious how the
components of that vector in any given basis are related to its values
at the critical points. However, as $Q$ is a ring, $f$ also acts by
multiplication on $Q$.  Since $Q$ is a vector space, and as
multiplication by $f$ is $\C$-linear, $f$ is also a {\em linear
  operator} on $Q$, which we can represent as a matrix $\mf{f}$ in our
basis of monomials in an obvious way. In particular, we could consider
$f=z_i$ as an operator or matrix $\mf{z_i}$ in this sense. Clearly,
the action of $\mf{z_i}$ on most basis vectors (for $d_i < d-2$) is
totally obvious,
\begin{equation}
\mf{z_i} : z_{1}^{d_1} \cdots z_i^{d_i} \cdots z_{N}^{d_N} \rightarrow
z_{1}^{d_1} \cdots z_i^{d_i+1} \cdots z_{N}^{d_N}, ~~~ d_i < d-2.
\end{equation}
However, for $d_i = d-2$ we must use the relations in the ideal to
re-express the resulting monomial as a linear combination of the basis
elements. Assuming that this is done for each $z_i$, we obtain $N$
matrices $\mf{z_i}$, each representing multiplication by one of the
$z_i$. We can consider the characteristic equation for each of these
matrices,
\begin{equation}
  \label{eq:chareq}
  0 = \det{(\mf{z_i} - \lambda \Bid)} = P_i(\lambda),
\end{equation}
which is a polynomial equation of degree $(d-1)^N$ in $\lambda$.
Generically, this equation will have $(d-1)^N$ distinct roots,
corresponding to the $(d-1)^N$ eigenvalues of the matrix $\mf{z_i}$. In
particular, the matrix $\mf{z_i}$ must be diagonal in the corresponding
basis of eigenvectors. Since $Q$ is a commutative ring, all
matrices corresponding to multiplication by functions must mutually
commute, and this must in particularly be true for the $N$ matrices
$\mf{z_i}$. Thus, all the matrices corresponding to multiplication by
functions $f \in \C[z_1,\ldots,z_N]$ can be simultaneously
diagonalized in the basis of eigenvectors of the $\mf{z_i}$. In particular,
in this eigenbasis, $Q$ splits {\em as a ring} into the direct sum of
$(d-1)^N$ rings isomorphic $\C$, 
\begin{equation}
  \label{eq:ringstruct}
  Q = \C[z_1, \ldots, z_N] / \< \del_i W \> \cong
  \underbrace{\C \oplus \cdots \oplus \C}_{(d-1)^N}.
\end{equation}
each of which one may naturally associate with the ring of functions
on one of the critical points of $W$.  Thus, generalizing our result
from the toy model, the eigenvalue of $\mf{f}$ associated with each
eigenvector corresponds to the value of $f$ at the corresponding
critical point. If we take $f=z_i$, then the eigenvalues of $\mf{z_i}$
are precisely the $z_i$ coordinates of the critical points. Further,
this means that the trace of $\mf{f}$ in {\em any basis} is the sum of
the values of $f$ at the critical points of $W$. In particular, we can
do this trace explicitly in the monomial basis. Now, as powers of $f$
correspond to powers of the corresponding matrix $\mf{f}$, we can take
their traces to compute the sums of powers of the values of $f$ at the
critical points as well. Thus, we see that by computing traces in the
monomial basis, we can compute all holomorphic moments of $f$ at the
critical points of a generic superpotential $W$. In other words, we
can effectively compute all (holomorphic) statistical properties of
the values of any polynomial $f$ at the critical points of $W$.

\subsubsection{Resultants and Genericity} \label{subsubsec:generic}

There are two basic ways in which the generic situation of $(d-1)^{N}$
vacua can break down. Roots of the system of polynomials can coincide,
or roots can run off to infinity. The former is something that already
was implicit in the setup of \cite{Arkani-Hamed:2005yv}, where the $N$
chiral fields were assumed to be decoupled. When roots run together,
the tunneling between the respective vacua is no longer suppressed,
and one should not really count them as independent vacua. It is not
difficult to modify the following considerations to deal with this
situation, but we will not discuss this further here (see \cite{CLO}
for details).  Having roots ``run off to infinity'' is a phenomenon
that can occur only when interactions between the fields are turned
on, and our analysis breaks down in this case. 

It turns out \cite{CLO} that both the collision and expulsion of roots
can be captured by a single genericity condition.\footnote{It is a
  certain integral polynomial in the coefficients $A_{\I_n}$ of $W$
  known as the u-resultant.} As it is only the latter non-genericity
which is fatal, it is useful to have a precise criterion when it
obtains. In the case of a quadratic superpotential, we saw that the
critical point runs off to infinity precisely when $\det C = 0$.  Just
as before, we can describe this more precisely by considering the
homogenization of the polynomial equations $\del_i W=0$ and lifting
them to to equations in $\CP{N}$. It turns out (see \cite{CLO} Chapter
3) that there is an analogous integral polynomial in the coefficients
of the monomials in $W$ of highest degree $A_{\I_d}$ known as a
multi-polynomial resultant which plays the same role as $\det C$ in
degree $>2$. The vanishing of this particular resultant indicates that
some solutions have ``run off to infinity'', or equivalently, the
presence of solutions in the compliment of the open set $U_0$. We will
refer to the vanishing locus of the resultant as the {\it
  discriminant locus}.  While we will not describe this resultant in
general, we will be able to compute it exactly in an example $d=N=3$
and see the advertised behavior explicitly in the next section.
Finally, note that the cubic polynomial superpotentials we are
interested in are approximations valid only within some small polydisk
about the origin. Indeed, in the real situation, we don't literally
have roots running off to infinity. Once they leave the polydisk we
are considering, our approximation of truncating to the renormalizable
terms in the Lagrangian breaks down.  We simply no longer {\it trust}
those solutions which have wandered too far from the origin.

\section{Holomorphic Moments and the Statistics of SUSY
  Vacua}\label{sec:Stats} 

We will now restrict our consideration to the physically relevant case of
a renormalizable, R-symmetric, cubic superpotential for $N$ chiral
fields in Fermat form,
\begin{equation}
  W(z_i) = \sum_{i=1}^N \left( \tfrac{1}{3} z_i^3 - a_i z_i \right) 
- \sum_{i < j < k}^N b_{ijk} z_i z_j z_k,
\end{equation}
where we recall that $b_{ijk}$ is symmetric with vanishing diagonals,
i.e. $b_{iii} = b_{iij} = b_{iji} = b_{jii} =0$. Dimensional analysis
as well as the $S_N\ltimes (\Z_3)^N$ symmetry will be important tools
for the analysis to follow. As the superpotential $W$ has mass dimension
$3$, the $z_i$ must have mass dimension $1$ while the $a_i$ have mass
dimension $2$ and the $b_{ijk}$ are dimensionless. Further,
$\boldsymbol{\zeta} \in (\Z_3)^N$ acts as,
\begin{equation}
 z_i \rightarrow z_i \zeta_i, ~~~~  a_i \rightarrow a_i \zeta_i^{-1},
 ~~~~ b_{ijk} \rightarrow b_{ijk} \zeta_i^{-1} \zeta_j^{-1}
 \zeta_k^{-1}, ~~~~ \boldsymbol{\zeta} = \{ \zeta_i \},
\end{equation}
where $\zeta_i$ is in the $i^{th}$ $\Z_3$, under which $z_i$ has
charge $+1$ while the $a_i$ and $b_{ijk}$ have charge $-1$.  Note that
the equations for the critical points of the cubic in Fermat form are,
\begin{equation}
  \label{eq:critp}
  z_i^2 - a_i - \sum_{j<k} b_{ijk} z_j z_k=0,
\end{equation}
and a basis for the quotient ring $Q$ is given by the $2^N$ monomials,
\begin{equation}
  \label{eq:basisN}
  \{ 1,z_{i}, z_{i} z_{j}, \ldots, z_1 \cdots z_N \},
\end{equation}
where at most one power of each $z_i$ appears in each monomial. The
residue of $W$ in $Q$ written in this basis is,
\begin{equation}
  \label{eq:Wres}
  W = -\frac{2}{3} \sum_{i=1}^N a_i z_i.
\end{equation}
Note that $\tr{\mf{W}}$ and all other holomorphic moments of odd
powers of $\mf{W}$ or $\mf{z_i}$ vanish due to the R-symmetry.
Explicitly, since the relations (\ref{eq:critp}) are even, any odd
power of the operator $\mf{z_i}$ maps monomials of even degree to
those of odd degree and vice-versa and therefore never has any
non-zero diagonal components.

We will be interested in determining the distribution of values of $W$
among the $2^{N}$ vacua which appear for generic values of the
couplings. These are the eigenvalues of $\mf{W}$.  By the R-symmetry,
the values of $W$ at the critical points occur in pairs, $\pm
\lambda$, and are the roots of the characteristic equation,\footnote{
  Note that in general, the number of roots that run off to infinity
  as one approaches the discriminant locus is controlled by the order
  of the pole of $F_{2^{N-1}}(a,b)$. For a pole of order $m$, $2m$
  roots run off to infinity. If the order of the pole is less than the
  maximal ($2^{N-1}$), then some of the roots remain finite, even in
  the limit.}
\begin{equation}
   0 = \det(\tfrac{3}{2}\mf{W} -\lambda \Bid) \equiv P(\lambda^{2}) =
   (\lambda^{2})^{2^{N-1}} - 2^{N-1}\sum_{k=1}^{2^{N-1}}
   F_{k}(a,b)(\lambda^{2})^{2^{N-1}-k} .
\end{equation}
We can express the $F_k (a,b)$ in terms of the holomorphic moments of
$\mf{W}$ by noting that,
\begin{equation}
   \det(\tfrac{3}{2}\mf{W} -\lambda \Bid) = \exp
   \tr{\log(\tfrac{3}{2}\mf{W} -\lambda \Bid)} = \lambda^{2^{N}}
   \exp{\tr{-\sum_{k=1}^{2^{N-1}} \frac{1}{2k} \left( \frac{3}{2}
   \frac{\mf{W}}{\lambda} \right)^{2k}} } ,
\end{equation}
where the last equality is only meaningful for positive powers of
$\lambda$ and encodes the form of the $F_k (a,b)$.  Dimensional
analysis and invariance under the $S_N\ltimes (\Z_3)^N$ symmetry can
be used to restrict the form of the $F_{k}(a,b)$. They must be
homogeneous polynomials in the $a_{i}$, of degree $3k$, whose
coefficients are rational functions of the $b_{ijk}$. Of particular
importance to us is $F_{1}$, which is proportional to the
``holomorphic variance'' of $\mf{W}$. Using the symmetries, we may
parameterize this variance as,
\begin{equation}\label{eq:F1def}
\vvev{\tfrac{9}{4}\mf{W}^2} \equiv 2^{-N} \tr{\tfrac{9}{4}\mf{W}^2} = F_{1}(a,b) = f(b)\sum_{i}a_{i}^{3} +\sum_{i<j<k}
 g^{ijk}(b)a_{i}a_{j}a_{k}+\sum_{i\neq j} h^{ij}(b) a_{i}^{2}a_{j},
\end{equation}
where the normalization factor converts the trace into the average
value of $\tfrac{9}{4} W^2$ among the $2^{N}$ critical points.  Now,
as $F_{1}$ must be invariant under $S_{N}\ltimes (\BZ_{3})^N$, we see
that $f(b)$ must also be invariant, while $g^{ijk}(b)$ is symmetric
with vanishing diagonals and must have charge $+1$ under the $i^{th}$,
$j^{th}$, and $k^{th}$ $\BZ_3$, and $h^{ij}(b)$ has vanishing diagonal
and must have charge $-1$ under the $i^{th}$ and $+1$ under the
$j^{th}$ $\BZ_3$. While we will not be able to compute the coefficient
functions $f(b)$, $g^{ijk}(b)$, and $h^{ij}(b)$ in closed form for
arbitrary large $N$, we can certainly do so for any small fixed $N$.
We will now turn to the simple example of the case $N=3$.

\subsection{Example I - $W$ odd, $d=3$, $N=3$} \label{subsec:Neq3}

Consider the case of the generic, odd, cubic superpotential in three
variables in Fermat form,
\begin{equation}
  \label{eq:wd}
  W= \frac{1}{3} \left( z_1^3 + z_2^3 + z_3^3 \right) -\left( a_1 z_1 + a_2
  z_2 + a_3 z_3 + b z_1 z_2 z_3 \right).
\end{equation}
The equations for its critical points are,
\begin{equation}
  \label{eq:wcrit}
  z_1^2 - a_1 - b z_2 z_3 = 0, ~~~~
  z_2^2 - a_2 - b z_1 z_3 = 0, ~~~~
  z_3^2 - a_3 - b z_1 z_2 = 0.
\end{equation}
A basis for the quotient ring $Q$ is given by the $2^3$ monomials,
\begin{equation}
  \label{eq:basismon}
  \{ 1, z_1, z_2, z_3, z_1 z_2, z_1 z_3, z_2 z_3, z_1 z_2 z_3 \}.
\end{equation}
In this basis, the residue of $W$ in $Q$ is,
\begin{equation}
  \label{eq:wdres}
  W= -\frac{2}{3} \left( a_1 z_1 + a_2
  z_2 + a_3 z_3 \right).
\end{equation}
It is an easy exercise to determine the form of the matrix $\mf{W}$,
\begin{equation}
\mf{W} = -\frac{2}{3} \left(
\begin{array}{cccccccc}
0 & {a_1}^2 & {a_2}^2 & {a_3}^2 & 0 & 0 & 0 &
  \frac{3 a_1 a_2 a_3 b}{1 - b^3} \\
a_1 & 0 & 0 & 0 & \frac{{a_2}^2 + a_1 a_3 b^2}{1 - b^3} & 
   \frac{{a_3}^2 + a_1 a_2 b^2}{1 - b^3} & \frac{2 a_2 a_3 b}{1 - b^3} & 0\\
a_2 & 0 & 0 & 0 & \frac{{a_1}^2 + a_2 a_3 b^2}{1 - b^3} & \frac{2 a_1 a_3
  b}{1 - b^3} &  \frac{{a_3}^2 + a_1 a_2 b^2}{1 - b^3} & 0\\
a_3 & 0 & 0 & 0 & \frac{2 a_1 a_2 b}{1 - b^3} & \frac{{a_1}^2 + a_2 a_3 b^2}{1 - b^3} & 
   \frac{{a_2}^2 + a_1 a_3 b^2}{1 - b^3} & 0\\
0 & a_2 & a_1 & a_3 b  & 0 & 0 & 0 & \frac{{a_3}^2 + 2 a_1 a_2
  b^2}{1 - b^3}\\ 
0 & a_3 & a_2 b  & a_1 & 0 & 0 & 0 & \frac{{a_2}^2 + 2 a_1 a_3 b^2}{1 - b^3}\\
0 & a_1 b  & a_3 & a_2 & 0 & 0 & 0 & \frac{{a_1}^2 + 2 a_2 a_3 b^2}{1 - b^3}\\
0 & 0 & 0 & 0 & a_3 & a_2 & a_1 & 0 
\end{array}
\right).
\end{equation}
Now, it is clear that $\tr{\mf{W}} = 0$ as expected from the
R-symmetry. More interesting is the fact that we can just as easily compute,
\begin{equation}
  \label{eq:trwsquared}
\vvev{ \tfrac{9}{4} \mf{W}^2} = 2^{-3}\tr{\tfrac{9}{4}\mf{W}^2} = 
      \left( \frac{(a_1^3 + a_2^3 + a_3^3) \left( 1 -  \frac{1}{4} b^3
      \right) + \frac{9}{2} a_1 a_2 a_3 b^2}{1 - b^3} \right),
\end{equation}
as well as all higher moments of $W$. In particular, we can explicitly
read off the coefficient functions of \eqref{eq:F1def} for the $N=3$ case,
\begin{equation}
\begin{split}
\label{eq:fghN3}
   f(b)&= \frac{1}{4}\frac{4-b^{3}}{1-b^{3}},\\
   g(b)&= \frac{9 b^{2}}{2(1-b^{3})},\\
   h(b)&= 0.
\end{split}
\end{equation}
Further, if we wish to explicitly compute the values of $W$ at the
critical points, we can use these moments to compute the
characteristic polynomial for $W$ and numerically solve it.
Indeed, this is easy to do using Mathematica, and one can verify
that numerical solution of the above equations through other means
agree with the eigenvalue methods. Further, using similar techniques,
we can also compute the moments of any polynomial over these vacua. 

Finally, we note that \eqref{eq:fghN3} are singular in the limit that
$b$ approaches a third root of unity. It is easy to check numerically
that six of the eight critical points run off to infinity in this
limit, and that these three points indeed comprise the discriminant
locus of the $N=3$ case. Thus, we see explicitly in this case that the
multi-polynomial resultant we discussed earlier must be proportional to
$(1-b^3)$ for $N=3$.

\subsection{Example II - The Fermat Cubic for Large $N$ and small $b_{ijk}$}
\label{subsec:LargeN}

Since the dimension of the vector space $Q$ grows exponentially with
the number of variables, the matrix methods introduced above become
quickly intractable, even numerically, for $N \gtrsim 10 $. In
particular, explicit computations require that we can compute
holomorphic moments like
\begin{equation}
  \label{eq:trWsq}
  \vvev{\tfrac{9}{4} \mf{W}^2} = 2^{-N} \tr{\tfrac{9}{4} \mf{W}^2} =
  \sum_{i,j} a_i a_j (2^{-N} \tr{\mf{z_i} \mf{z_j}}) = \sum_{i,j} a_i
  a_j \vvev{\mf{z_i} \mf{z_j}} ,
\end{equation}
which in turn require that we can compute the moments of products of
the $\mf{z_i}$'s. This could be done if we could easily compute the
matrix elements of $\mf{z_i}$ in this basis. However, as we mentioned
earlier, this is very computationally intensive, and involves the
inversion of exponentially large matrices.

Thus, it is useful to understand if there exists a limit in which the
above methods become more tractable at large $N$, at least
perturbatively.  Now, note that the decoupled limit with $b_{ijk}=0$
considered in \cite{Arkani-Hamed:2005yv} is certainly such a simple
case, so the $b_{ijk}$ seem to be natural candidates for small
parameters. Indeed, radiative stability of the K\"{a}hler potential
requires that generically $|A_{ijk}| \sim \mcO(N^{-1})$ while at most
$\mcO(N)$ of them may be $\mcO(1)$. That is, the $b_{ijk}$'s are {\it
  generically small}. There are, of course, a large number of them, so
this large-$N$ suppression of the magnitudes of the $b_{ijk}$ is
compensated by the sum. But say that we make the further assumption
that perturbation theory is good --- that is, that the magnitude of
the $b_{ijk}$ is further suppressed by some small parameter $\delta$.
Then it makes sense to expand our expressions for $\vvev{f(\mf{z})}$
as a (formal) power series in the $b_{ijk}$.

\subsubsection{Computations of Holomorphic Moments for Small $b_{ijk}$}

To begin with, we can use dimensional analysis as well as
the $S_N\ltimes (\Z_3)^N$ symmetry to constrain the form of successive
terms in the power series expansions for $\vvev{f(\mf{z})}$. Let us
consider $\vvev{\mf{z_i} \mf{z_j}}$ as an example. First note that if we
have $i=j$,
\begin{equation}
  \label{eq:trzsq}
  \vvev{\mf{z_i}^2} = \vvev{a_i \Bid + \sum_{j<k} b_{ijk} \mf{z_j}
  \mf{z_k}} = a_i + \sum_{j<k} b_{ijk} \vvev{\mf{z_j} \mf{z_k}},
\end{equation}
so we only need focus on the case $i \neq j$. Using the symmetries and
dimensional analysis, it is easy to check that the first few terms in
the (formal) power series expansion of $\vvev{\mf{z_i}\mf{z_j}}$ for
$i \neq j$ must take the form, 
\begin{equation}
  \label{eq:trztwo}
  \vvev{\mf{z_i} \mf{z_j}} = \a^{(2)}_1 \sum_k b_{ijk}^2 a_k + \sum_{k<l}
        b_{ikl}b_{jkl} \left[ \a^{(3)}_1 ( b_{jkl}a_i + b_{ikl}a_j ) +
        \a^{(3)}_2 ( b_{ijk} a_l + b_{ijl} a_k ) \right] +\CO(b^{4}),
\end{equation}
where the $\a^{(n)}_i \in \Q$ are rational numbers which we will see
are {\em independent} of $N$ for any fixed $\a^{(n)}_i$ and $N$
sufficiently large. The $\a^{(n)}_i$ can be found by computing the
diagonal matrix elements of the operator $\mf{z_i} \mf{z_j}$ in the
monomial basis, summing them, and dividing by $2^N$. Of course, the
only way one obtains a diagonal element is if the relations
\eqref{eq:critp} are used, so such diagonal elements only arise from
the action of $\mf{z_i} \mf{z_j}$ on basis elements containing $z_i$
or $z_j$. Each use of the relation results either in the introduction
of one factor of $a_i$ or $b_{ijk}$, so the $\a^{(n)}_i$ can be
computed using the relations precisely $(n+1)$ times. For example, the
contribution of the monomial $z_i$ to $\a^{(n)}_1$ is found by,
\begin{equation}
\begin{split}
  \label{eq:matelmts}
  \mf{z_i} \mf{z_j} \cdot z_i & =  a_i z_j + \sum_{k < l \neq j}
  b_{ikl} z_k z_l z_j +  \sum_{l \neq (i,j)} b_{ijl} \mf{z_j} \mf{z_l}
  \cdot z_j  \\ 
& = \cdots + \sum_{l \neq
  (i,j)} b_{ijl} \Bigl( a_j z_l + \sum_{m < n \neq (l,j)}
  b_{jmn} z_m z_n z_l +  \sum_{m \neq (j,l)} b_{jlm} \mf{z_l} \mf{z_m}
  \cdot z_l \Bigr) \\
& = \cdots + \sum_{l \neq (i,j)} \sum_{m \neq (l,j)} b_{ijl} b_{jlm} a_l
  z_m + \mcO(b^3) \\
& = \cdots + \sum_k b_{ijk}^2 a_k z_i + \cdots.
\end{split}
\end{equation}
In fact, we can greatly simplify this computation by noting that the
$S_N$ symmetry implies that we only need to compute how many times any
\emph{single} term of a sum over dummy indices appears in the diagonal
matrix elements. For example, to find $\a^{(2)}_1$, we only need to
ask how many times the single term $b_{ijk}^2 a_k$ with $i$,$j$, and
$k$ all fixed appears on the diagonal of $\mf{z_i} \mf{z_j}$. This is
very easy to do. First of all, this term only involves $a_k$ and one
particular coefficient $b_{ijk}$. As these two coefficients arise in
the relations \eqref{eq:critp} only in terms involving $z_i$, $z_j$,
and $z_k$, we can completely ignore all the other variables, which
just go along for the ride.\footnote{ Formally, this is equivalent to
  working in $Q$ modulo the ideal $\langle z_l \rangle$, $l \neq
  i,j,k$.} Thus, to calculate $\a^{(2)}_1$, we only need to compute
the coefficient of the term $b_{ijk}^2 a_k$ in the trace of $\mf{z_i}
\mf{z_j}$ restricted to the $2^3$ monomials of the form $z_i^{a_1} z_j^{a_2}
z_k^{a_3}$, $a_i = 0,1$ in our basis \eqref{eq:basisN}, ignoring all
terms involving any other variables (and divide the result by $2^{3}$).
That is, we have shown that this coefficient can be computed by
considering just the case $N=3$. For instance, we can expand
\eqref{eq:trwsquared} in a power series in $b$ and find $\a^{(2)}_1 =
\tfrac{3}{4}$.

We can more easily compute this using a diagrammatic technique.  To
compute the contribution of $\mf{z_i} \mf{z_j} \cdot z_i^{a_1}
z_j^{a_2} z_k^{a_3}$ to $\a^{(2)}_1$, first draw a labeled external
line for $\mf{z_i}$ and $\mf{z_j}$ as well as all the $z_n$'s present
in the basis element.  Each $b_{ijk}$ corresponds to a vertex which
takes in two identically labeled lines corresponding to one of its
indices and outputs two lines corresponding to the other two indices,
while $a_k$ is an external sink for a pair of $z_k$ lines. Construct
all possible graphs with two $b_{ijk}$ vertices, and one $a_k$
external sink from the given lines, noting that graphs which differ by
a choice of which lines of the same index are contracted at a vertex
or sink are equivalent. Count the number of distinct graphs you drew
and divide by $2^3$ - this is the contribution. It turns out that
there is precisely one possible graph for each basis element with
either $z_i$ or $z_j$ in it, so we again find $\a^{(2)}_1 =
\tfrac{3}{4}$.

The above method, of course, generalizes in the obvious way to the
computation of any $\a^{(n)}_i$ by restricting ourselves to just the
relevant $N=n+1$ variables. For example, to compute the coefficient
$\a^{(3)}_1$ of $b_{ikl} b_{jkl}^2 a_i$, we consider $N=4$ and graphs
with one $b_{ikl}$ vertex and two $b_{jkl}$ vertices as well as a
single sink $a_i$. Note that it is useful to start with monomials of
lowest total degree and proceed to those of higher degree as any graph
associated with a given basis monomial also contributes to all
monomials it divides. Thus, without much trouble, one can compute the
terms cubic in $b_{ijk}$,
\begin{equation}
\vvev{\mf{z_{i}}\mf{z_{j}}}= \tfrac{3}{4} \Bigl[  \sum_k b_{ijk}^2 a_k
     + \sum_{k<l} b_{ikl}b_{jkl} \left[
          (b_{jkl}a_i + b_{ikl}a_j) + 2( b_{ijk} a_l + b_{ijl} a_k )
        \right] +\CO(b^{4})
  \Bigr],
\end{equation}
so $\a^{(3)}_1 = \tfrac{3}{4}$, $\a^{(3)}_2 = \tfrac{3}{2}$. Using
this result as well as \eqref{eq:trzsq} we can compute,
\begin{equation}
\begin{split}
  \label{eq:Wsqhol}
  \vvev{\tfrac{9}{4}\mf{W}^2} & = \left( \sum_i a_i^2
  \vvev{\mf{z_i}^2} + 2 \sum_{i<j} a_i a_j \vvev{\mf{z_i} \mf{z_j}}
  \right)
   = \sum_i a_i^3  + \sum_{i<j} \left( 2 a_i a_j + \sum_{l}
  a_l^2 b_{ijl} \right) \vvev{\mf{z_i} \mf{z_j}} \\ 
& = \left(1+\tfrac{3}{4}\sum_{j<k<l}b_{jkl}^{3}\right) \sum_i a_i^3 
+ \sum_{i<j<k} \tfrac{9}{2} \left(
  b_{ijk}^2+2\sum_{l}b_{ijl}b_{jkl}b_{ikl} \right) a_i a_j a_k \\
  &\qquad + \sum_{i\neq j} \left( \tfrac{3}{4}
  \sum_{k<l}b_{ikl}b_{jkl}^{2} \right)  a_{i}^{2}a_{j} + \mcO(b^4).
\end{split}
\end{equation}
Thus, we can read off the coefficients functions of
\eqref{eq:F1def} up to terms of order $b^4$, 
\begin{equation}
\begin{split}
  \label{eq:fgh}
f(b) &=1+\tfrac{3}{4}\sum_{j<k<l}b_{jkl}^{3}+\dots, \\ 
g^{ijk}(b) &=  \tfrac{9}{2} \left(
  b_{ijk}^2+2\sum_{l}b_{ijl}b_{jkl}b_{ikl} \right) +\dots, \\
h^{ij}(b) &= \tfrac{3}{4} \sum_{k<l}b_{ikl}b_{jkl}^{2}+\dots.
\end{split}
\end{equation}
As noted, each additional power of $b$ brings an additional sum over a
dummy index, which has $\CO(N)$ terms. If $|b_{ijk}| \sim N^{-1}\d$,
then the term of order $b^{k}$ goes like $N \d^{k}$, so this is a
systematic expansion in powers of $\d$. 

We can continue and compute higher holomorphic moments in an expansion
in powers of $b_{ijk}$ - for instance, the following holomorphic
moments will be useful (with $i \neq j \neq k \neq l$),
\begin{equation}\label{eq:quarticmoments}
\begin{split}
   \vvev{\mf{z_{i}}^{4}}-\vvev{\mf{z_{i}}^{2}}^{2} &=
    \sum_{j<k}b_{ijk}^{2}a_{j}a_{k}+\CO(b^{3})\\ 
    \vvev{\mf{z_{i}}^{3}\mf{z_{j}}}
        -\vvev{\mf{z_{i}}^{2}}\vvev{\mf{z_{i}}\mf{z_{j}}} &=
    \sum_{k}b_{ijk}^{2}a_{i}a_{k}+\CO(b^{3})\\ 
    \vvev{\mf{z_{i}}^{2}\mf{z_{j}}^{2}}
        -\vvev{\mf{z_{i}}^{2}}\vvev{\mf{z_{j}}^{2}} &=
    \sum_{k<l}b_{ikl}b_{jkl}a_{k}a_{l} +\CO(b^{3})\\ 
    \vvev{\mf{z_{i}}^{2}\mf{z_{j}}^{2}}
        -\vvev{\mf{z_{i}}\mf{z_{j}}}^{2} &= a_{i} a_{j} +
    \sum_{k<l}b_{ikl}b_{jkl}a_{k}a_{l} +\CO(b^{3}) \\ 
    \vvev{\mf{z_{i}}^{2}\mf{z_{j}}\mf{z_{k}}}
        -\vvev{\mf{z_{i}}^{2}}\vvev{\mf{z_{j}}\mf{z_{k}}} &=
    b_{ijk}a_{j}a_{k} +\CO(b^{3})\\ 
    \vvev{\mf{z_{i}}^{2}\mf{z_{j}}\mf{z_{k}}}
        -\vvev{\mf{z_{i}}\mf{z_{j}}}\vvev{\mf{z_{i}}\mf{z_{k}}} &=
    b_{ijk}a_{j}a_{k} +
    \tfrac{3}{4}\sum_{l}b_{jkl}^{2}a_{i}a_{l}+\CO(b^{3})\\ 
    \vvev{\mf{z_{i}}\mf{z_{j}}\mf{z_{k}}\mf{z_{l}}}
        -\vvev{\mf{z_{i}}\mf{z_{j}}}\vvev{\mf{z_{k}}\mf{z_{l}}} &= \CO(b^{3}).
\end{split}
\end{equation}
Therefore, we see that it is possible to effectively compute
holomorphic moments and any desired holomorphic statistical property
in the limit of large $N$ as a systematic expansion in small
$b_{ijk}$. 

For any \emph{fixed} $N$, we can, with more effort, crank out some
explicit formul\ae, capturing the dependence on the discriminant
locus. For $N=3$, we see explicitly, that each of the moments in
\eqref{eq:quarticmoments} has a double pole at the discriminant locus.
\begin{equation}\label{eq:QuarticNeq3Moments}
\begin{split}
   \vvev{\mf{z_{1}}^{4}}-\vvev{\mf{z_{1}}^{2}}^{2} &=
   \frac{ b^{2}(8(2+b^{3})a_{2}a_{3}+3 b^{4}a_{1}^{2})}{16(1-b^{3})^{2}}\\ 
    \vvev{\mf{z_{1}}^{3}\mf{z_{2}}}
        -\vvev{\mf{z_{1}}^{2}}\vvev{\mf{z_{1}}\mf{z_{2}}} &=
    \frac{b^{2}((16-b^{3})a_{1}a_{3}+12b a_{2}^{2})}{16(1-b^{3})^{2}}\\ 
    \vvev{\mf{z_{1}}^{2}\mf{z_{2}}^{2}}
        -\vvev{\mf{z_{1}}^{2}}\vvev{\mf{z_{2}}^{2}} &=
    \frac{b^{3}((16-b^{3})a_{1}a_{2}+12b a_{3}^{2})}{16(1-b^{3})^{2}}\\ 
   \vvev{\mf{z_{1}}^{2}\mf{z_{2}}^{2}}-\vvev{\mf{z_{1}}\mf{z_{2}}}^{2}&=
   \frac{8(2+b^{3})a_{1}a_{2}+3 b^{4}a_{3}^{2}}{16(1-b^{3})^{2}}\\ 
    \vvev{\mf{z_{1}}^{2}\mf{z_{2}}\mf{z_{3}}}
        -\vvev{\mf{z_{1}}^{2}}\vvev{\mf{z_{2}}\mf{z_{3}}} &=
    \frac{b(8(2+b^{3})a_{2}a_{3}+3b^{4} a_{1}^{2})}{16(1-b^{3})^{2}}\\ 
    \vvev{\mf{z_{1}}^{2}\mf{z_{2}}\mf{z_{3}}}
        -\vvev{\mf{z_{1}}\mf{z_{2}}}\vvev{\mf{z_{1}}\mf{z_{3}}} &=
    \frac{b((16-b^{3})a_{2}a_{3}+12b a_{1}^{2})}{16(1-b^{3})^{2}}\\ 
\end{split}
\end{equation}

\section{Holomorphic Moments and Scanning}
\label{sec:Scanning}

Now, let us consider the relationship between the holomorphic moments
we have computed and the {\em real-valued} statistical properties of
the vacua.  $W=W_{1}+i W_{2}$ is complex. Averaged over the $2^{N}$
vacua, the mean value is, of course, zero.
\begin{equation*}
\vvev{W_{1}}=\vvev{W_{2}}=0
\end{equation*}
The variances involve the sums of squares of the the eigenvalues, and
so are encoded in $\tfrac{4}{9} 2 F_1 = \vvev{\mf{W}^2}$. The variances
we are interested in are $\delta W_{1}^{2}$, $\delta W_{2}^{2}$ and
the covariance, $\vvev{W_{1}W_{2}}$. Two linear combinations of these
are ``holomorphic'' in the eigenvalues,
\begin{subequations}\label{eq:Wvariances}
\begin{align}
   \delta W_{1}^{2}-\delta W_{2}^{2}&= \Real{\vvev{\mf{W}^2}}\\
   \vvev{W_{1}W_{2}}&= \Imag{\vvev{\mf{W}^2}}
\end{align}
The remaining linear combination requires more detailed knowledge,
though it is easy to find a lower bound from the Cauchy-Schwarz
inequality,
\begin{equation}
  \delta W_{1}^{2}+\delta W_{2}^{2} \geq |\vvev{\mf{W}^2}|
\end{equation}
\end{subequations}
In the following, we will turn to the implications of these
formul\ae\ when one averages over some ensemble of couplings. A lower
bound, like (\ref{eq:Wvariances}c) will be quite sufficient to prove
that a given coupling ``scans''. But one requires something of an {\it
  upper} bound in order to prove that a coupling {\it doesn't} scan.
These methods generalize straightforwardly to other coupling which
depend holomorphically on the $\phi_{i}$ (superpotential couplings for
the Standard model fields, holomorphic gauge couplings, {\it etc.}).

\subsection{Ensembles of Theories and the Scanning of $\Lambda$}

Let us return to \eqref{eq:F1def},  
\begin{equation}
  \label{eq:VarW} 
\vvev{\mf{W}^2} = \tfrac{4}{9} \Bigl[ f(b)\sum_{i}a_{i}^{3} +\sum_{i<j<k}
 g^{ijk}(b)a_{i}a_{j}a_{k}+\sum_{i\neq j} h^{ij}(b) a_{i}^{2}a_{j} \Bigr],
\end{equation}
which encodes the variance of $W$ among the $2^{N}$ vacua for fixed
values of the couplings. What we would like to do now is discuss what
happens if we have some ensemble of theories. For example, rather than
studying F-theory vacua for {\it fixed} flux, we might wish to study
the ensemble of all possible fluxes.

No one, currently, has a compelling proposal for what probability
measure to choose for that ensemble. Neither do we. What we hope to do
in this section is to explore the implications one can extract from
such a choice, given the results of the previous sections.

The authors of \cite{Arkani-Hamed:2005yv} worked with $b_{ijk}\equiv
0$, and assumed that the $a_{i}$ were $N$ independent random
variables, chosen from some common distribution. In their limit, the
second two terms of \eqref{eq:VarW} vanish, and we have
\begin{equation*}
\vvev{\mf{W}^2} =\tfrac{4}{9} \sum_{i=1}^{N}a_{i}^{3}
\end{equation*}
Since there are $N$ terms, and each is an independent random variable,
the sum grows like $\CO(N)$. At first glance, for nonzero $b_{ijk}$,
the situation appears to change dramatically. The second term in
\eqref{eq:VarW} contains $\CO(N^{3})$, and the third term contains
$\CO(N^{2})$ independent random variables. At least, that would be the
case if the $b_{ijk}$ coefficients were all generically of $\CO(1)$.
However, as we have argued, radiative stability requires that the
$b_{ijk}$ be generically small.  So, despite appearances, each of the
terms in \eqref{eq:F1def} is $\CO(N)$. The cosmological constant scans
(and, in fact, the variance grows like $\CO(N)$) in this more general
context, just as it did in the model of \cite{Arkani-Hamed:2005yv}
where $b_{ijk}\equiv 0$.

\subsection{Other Couplings}

Other holomorphic couplings which depend on the $\phi_{i}$ can be
treated similarly. Consider some such coupling, $c(\phi)$, which might
be a holomorphic gauge coupling, or a coupling in the superpotential
for the Standard Model fields. It is reasonable to assume that
$c(\phi)$ has definite parity under the $\BZ_{4}$ R-symmetry that
constrained the form of $W(\phi)$.  The crucial distinction in
\cite{Arkani-Hamed:2005yv} is between those cases where $c(\phi)$ is
odd and hence $\vvev{c(\phi)}=0$ (like the superpotential, $W$) versus
those for which $c(\phi)$ is even. The former couplings ``scan,''
whereas the latter do not: the standard deviation of $c(\phi)$ among
the $2^{N}$ vacua is much smaller that its mean value.

As with the superpotential, we assume that $c(\phi)$ has a Taylor
expansion, convergent in a polydisk of radius $M_{c}$. For simplicity,
we will take it to be dimensionless; the case of the $\mu$ parameter
in the Standard Model is an easy generalization. So we have,
\begin{equation}\label{e:cexp}
\begin{split}
  c_{\mathrm{ev}}(\phi) &= c_{0} + \sum_{i\leq j}
  c_{ij}\phi_{i}\phi_{j}/M_{c}^{2}+\dots\\
  c_{\mathrm{odd}}(\phi) &= \sum_{i} c_{i}\phi_{i}/M_{c} +\dots
\end{split}
\end{equation}
depending on the parity of $c(\phi)$.  As in our previous discussion,
radiative stability constrains the form of the coefficients in this
expansion for large-$N$. The $c_{ijk\dots}$ must be {\it generically
  small}. Specifically, we have constraints of the form
\begin{equation}
\begin{split}
   g^{i\overline{\imath}}g^{j \overline{\jmath}}c_{ij}
   \overline{c}_{\overline{\imath} \overline{\jmath}}&\sim \CO(N)\\ 
   \det{}_{i \overline {\imath} }( g^{j \overline{\jmath} } c_{ij}
   \overline{c}_{\overline{\imath}\overline{\jmath}}) & \sim \CO(1)
\end{split}
\end{equation}
and so forth for the higher coefficients. Thus, the mean value of the
$(2k)^{th}$ term in the series \eqref{e:cexp} (the mean value of the
odd terms vanish), rather than going like $N^{2k} (M_{r}/M_{c})^{2k}$
actually goes like $N (M_{r}/M_{c})^{2k}$. So, we have a systematic
expansion in powers of $\epsilon^{2}=(M_{r}/M_{c})^2$. For
sufficiently small $\epsilon$, the leading, $\epsilon^{2}$ term can
compensate for the overall factor of $N$, while the subleading terms
are negligible. More explicitly, if we let $a_i = M_r^2 \tilde{a}_i$
we have,
\begin{subequations}\label{eq:cvev}
\begin{gather}
    \vvev{c_{\mathrm{odd}}}=0\\
    \vvev{c_{\mathrm{ev}}}= c_{0}+ \epsilon^{2}
    \left[f(b)\sum_{i}c_{ii}\tilde{a}_{i}+ \sum_{i\neq j}h^{ij}(b)c_{ii}\tilde{a}_{j}+
    \frac{1}{2}\sum_{i\neq j\neq k} g^{ijk}(b) c_{ij}\tilde{a}_{k}\right]
    +\CO(\epsilon^{4})
\end{gather}
\end{subequations}
where the $\CO(\epsilon^{4})$ terms represent the contributions of
quartic and higher terms in $c(\phi)$. Note that the rational
functions of $b$: $f(b)$, $g^{ijk}(b)$ and $h^{ij}(b)$ are the
\emph{same} ones from \eqref{eq:F1def} that appeared in the
computation of the variances of the superpotential and were computed
up to terms of $\mcO(b^4)$ in \eqref{eq:fgh}. It is easy to check
that, as announced, the leading term behaves as $N\epsilon^{2}$, with
the subleading terms suppressed by higher powers of $\epsilon^{2}$.

The variance is calculated similarly. In the odd case,
\begin{equation}
   \vvev{c_{\mathrm{odd}}^{2}} = \epsilon^{2}
   \left[f(b)\sum_{i}c_{i}^{2}\tilde{a}_{i}+ \sum_{i\neq
   j}h^{ij}(b)c_{i}^{2}\tilde{a}_{j}+ \frac{1}{2}\sum_{i\neq j\neq k}
   g^{ijk}(b) c_{i}c_{j}\tilde{a}_{k}\right] +\CO(\epsilon^{4}).
\end{equation}
Thus, given a distribution for the couplings we can definitively show
that this coupling scans if this variance is non-vanishing.

In the even case, let $\hat{c} = c - \vvev{c}$. The variance
is then given by,
\begin{equation}\label{eq:cvar}
  \vvev{\hat{c}^{2}} = \sum_{i,j,k,l}
  \frac{c_{ij}c_{kl}}{M_c^4}\left[ \vvev{\mf{z_{i}} \mf{z_{j}}
  \mf{z_{k}} \mf{z_{l}}}
  - \vvev{\mf{z_{i}}\mf{z_{j}}}\vvev{\mf{z_{k}}\mf{z_{l}}} \right]
  +\CO(\epsilon^{6}).
\end{equation}
As in \S\ref{subsec:LargeN}, we can easily compute \eqref{eq:cvar} in
an expansion in powers of $b_{ijk}$ using \eqref{eq:quarticmoments}.
For finite $N$, we can do better -- for $N=3$, we have
\eqref{eq:QuarticNeq3Moments}. As with the superpotential, we can
write $\hat{c} = c_{1}+ic_{2}$, and we have
\begin{subequations}\label{eq:cvars}
\begin{align}
 \delta c_{1}^{2} - \delta c_{2}^{2} &= \Real \vvev{\hat{c}^{2}},\\
 \vvev{c_{1}c_{2}}&= \Imag \vvev{\hat{c}^{2}},\\
 \delta c_{1}^{2} +\delta c_{2}^{2} &\geq |\vvev{\hat{c}^{2}}|.
\end{align}
\end{subequations}
The right-hand-side of (\ref{eq:cvev}b) goes like $N\epsilon^{2}$,
whereas the right-hand-side of \eqref{eq:cvar} goes like
$N\epsilon^{4}$.  We would like to conclude that the standard
deviations, $\delta c_{1}/c_{1}$ and $\delta c_{2}/c_{2}$, behave as
$\tfrac{\sqrt{N}\epsilon^{2}}{N\epsilon^{2}}=\tfrac{1}{\sqrt{N}}$.

Unfortunately, (\ref{eq:cvars}c) is only a \emph{lower bound}, so more
detailed information is necessary to \emph{really show} that this
coupling does \emph{not} scan. Indeed, this is a potentially serious drawback to the whole notion of a ``friendly landscape.'' If the lower bound in (\ref{eq:cvars}c) drastically underestimates the true variance, then these couplings \emph{will} vary appreciably over this ensemble of vacua. And anthropic arguments, based on holding them fixed while varying other coupling, like the cosmological constant, are incorrect. 

\section{Generalizations}\label{sec:Generalizations}

The ``space'' of vacua discussed here is the \emph{complex} affine
algebraic variety, $\C[z_1, \ldots, z_N] / \< \del_i W \>$. As such,
it was amenable to the techniques of complex algebraic geometry. If we
were studying $N$ real scalar fields, $\phi_{i}$, with potential,
$V(\phi)$, we would be faced with a problem in \emph{real} algebraic
geometry. This problem is harder, both because real algebraic geometry
is harder, and less well-developed than the complex case and because
the characterization of the desired space of vacua is more subtle. We
are not interested in $\BR[x_1, \ldots, x_N] / \< \del_i V \>$. We are
only interested in \emph{minima} of $V$, as opposed to all critical
points. At large $N$, ``most'' critical points of $V$ are actually
saddle points, and it's algebraically a little awkward to pick out
just the minima.

A more interesting generalization is the case in which some of our
complex chiral multiplets are charged under a $U(1)^{k}$ gauge
symmetry. Physically, this leaves open the possibility of
supersymmetry-breaking, in this ``moduli'' sector, via the inclusion
of Fayet-Iliopoulos terms. Mathematically, this gauging moves us from
the realm of complex affine algebraic algebraic geometry to that of
\emph{toric} geometry. The toric version of our problem is nearly as
well-developed mathematically as the affine case we have discussed. It
would be very interesting to generalize our considerations to that
case.

Finally, we have deliberately eschewed discussion of the microphysics
that determines the (ensemble of) couplings in our low-energy
effective Lagrangian. Douglas and collaborators \cite{ Douglas:2003um,
  Ashok:2003gk, Douglas:2004kp, Douglas:2004zu, Denef:2004ze,
  Douglas:2004qg, Douglas:2004kc, Douglas:2004zg, Denef:2004cf,
  Douglas:2005df}, for instance, have pursued the idea of treating all
possible choices of fluxes in an F-theory compactification
``democratically,'' assigning equal weight to the low-energy theory
that arises from each choice of flux. Further developments along this
vein appear in \cite{Giryavets:2004zr, DeWolfe:2004ns,
  Blumenhagen:2004xx, Misra:2004ky, Conlon:2004ds, Kumar:2004pv,
  Acharya:2005ez}.  It has been argued \cite{DeWolfe:2005uu} that in
the context of type IIA flux compactifications this may not be a
reasonable choice, as there are instances where the number of possible
choices of flux is infinite. Independent of this more subtle question,
we believe that our methods will prove useful in analyzing the vacuum
structure of any give theory \emph{whenever} one has a large number,
$N$, of light chiral multiplets.

\section{Acknowledgments}

We would like to thank N. Arkani-Hamed, A. Bergman, G. Farkas, W.
Fischler, S. Keel, S. Paban, B. Sturmfels, J. Vaaler, and C. Vafa for
discussions. JD would like to thank the Harvard Theory Group and the
Perimeter Institute for hospitality while this work was in progress.

This material is based upon work supported by the National Science
Foundation under Grant Nos. PHY-0071512 and PHY-0455649, and by grant
support from the US Navy, Office of Naval Research, Grant Nos.
N00014-03-1-0639 and N00014-04-1-0336, Quantum Optics Initiative.


\bibliographystyle{utphys}
\bibliography{poly}
\end{document}